\documentclass[twocolumn,showkeys,showpacs,preprintnumbers,prd,superscriptaddress,nofootinbib]{revtex4-2}
\usepackage{graphicx} 
\DeclareUnicodeCharacter{2212}{\ensuremath{-}}

\pdfoutput=1

\usepackage[utf8]{inputenc}
\usepackage{amsmath,mathtools}

\usepackage{tensor}
\usepackage{amssymb}
\usepackage{physics}
\usepackage[colorlinks=true,linkcolor=Blue,citecolor=YellowOrange,urlcolor=magenta]{hyperref}

\usepackage[capitalize]{cleveref}
\usepackage{tabularx}
\usepackage{multirow}
\usepackage{float} 
\usepackage{lipsum}  
\usepackage{rotating}
\usepackage{bigints}
\usepackage{verbatim}

\usepackage{braket}
\usepackage{cancel}
\usepackage{makecell}
\usepackage{booktabs}
\usepackage[dvipsnames]{xcolor}
\usepackage{graphicx} 
\usepackage[caption=false]{subfig}
\usepackage{array}
\newcolumntype{C}{>{\centering\arraybackslash}X}
\usepackage{makecell}

\newcommand{\roughly}[1]{\mathrel{\raise.3ex\hbox{$#1$\kern-0.85em
\lower1ex\hbox{$\sim$}}}}

\newcolumntype{C}{>{\centering\arraybackslash}X}

\begin{document}

\title{Dynamical Dark Energy Meets Varying Electron Mass: Implications for Phantom Crossing and the Hubble Constant}

\author{Adam Smith}
\email{asmith69@sheffield.ac.uk}
\affiliation{School of Mathematical and Physical Sciences, University of Sheffield, Hounsfield Road, Sheffield S3 7RH, United Kingdom} 

\author{Emre \"{O}z\"{u}lker}
\email{e.ozulker@sheffield.ac.uk}
\affiliation{School of Mathematical and Physical Sciences, University of Sheffield, Hounsfield Road, Sheffield S3 7RH, United Kingdom} 

\author{Eleonora Di Valentino}
\email{e.divalentino@sheffield.ac.uk}
\affiliation{School of Mathematical and Physical Sciences, University of Sheffield, Hounsfield Road, Sheffield S3 7RH, United Kingdom} 

\author{Carsten van de Bruck}
\email{c.vandebruck@sheffield.ac.uk}
\affiliation{School of Mathematical and Physical Sciences, University of Sheffield, Hounsfield Road, Sheffield S3 7RH, United Kingdom} 

\date{\today}

\begin{abstract}
We investigate the interplay between varying electron mass ($m_e$) and dynamical dark energy by analysing the Chevallier-Polarski-Linder (CPL) parametrization and its non-crossing variants, both with and without a varying-$m_e$ component. Our aim is to assess whether the preference for late-time dynamics and phantom divide line (PDL) crossing persists when early-time physics is introduced, and whether these combined models improve the alleviation of the Hubble tension compared to the varying-$m_e$ extension alone. Using the latest CMB, BAO, and supernova datasets, we derive updated constraints on $\Lambda$CDM, CPL, and their extensions, and examine their impact on $H_0$ and the preference for late-time dynamics. We find that $\Lambda$CDM+$m_e$ yields the largest upward shift in $H_0$, while replacing $\Lambda$ with the CPL parametrization or its non-crossing variants provides modest improvements in the overall fit. The data consistently favour dynamical dark energy and a phantom divide line crossing at scale factors $a_{\rm c}\simeq0.6–0.9$, and these preferences remain robust, though somewhat weaker ($\gtrsim2\sigma$), when the electron mass is also allowed to vary. Among the late-time models, CPL performs better than its non-crossing variants, further reinforcing the evidence for a genuine phantom divide crossing. The alleviation of the $H_0$ tension in the varying-$m_e$ case arises from late-time data breaking the strong $\Omega_m$–$m_e$ degeneracy in the CMB, while the additional degrees of freedom in CPL models allow the late-time dynamics to absorb this impact, thereby weakening the degeneracy breaking and further lowering $H_0$ through their ability to yield a decreasing dark energy contribution.
\end{abstract}

\maketitle

\section{Introduction}
The persistent discrepancy between the value of the Hubble constant $H_0$ inferred from the cosmic microwave background (CMB) within $\Lambda$CDM and that measured by late-time distance-ladder methods remains one of the most significant challenges in modern cosmology~\cite{Verde:2019ivm,DiValentino:2020zio,DiValentino:2021izs,Perivolaropoulos:2021jda,Schoneberg:2021qvd,Shah:2021onj,Abdalla:2022yfr,DiValentino:2022fjm,Kamionkowski:2022pkx,Giare:2023xoc,Hu:2023jqc,Verde:2023lmm,DiValentino:2024yew,CosmoVerse:2025txj}. While $\Lambda$CDM anchored on \textit{Planck} 2018 data yields $H_0 = 67 \pm 0.5$\,km\,s$^{-1}$\,Mpc$^{-1}$~\cite{Planck:2018vyg}, the combination of the independent CMB datasets from SPT+ACT provides $H_0 = 66.59 \pm 0.46$\,km\,s$^{-1}$\,Mpc$^{-1}$~\cite{SPT-3G:2025bzu}. On the distance-ladder side, the original SH0ES calibration of supernova distances with Cepheids yielded $H_0 = 73.04 \pm 1.04$\,km\,s$^{-1}$\,Mpc$^{-1}$~\cite{Riess:2021jrx}, establishing a $>5\sigma$ tension. The latest SH0ES determination, including an additional geometric anchor, reports $H_0 = 73.17 \pm 0.86$\,km\,s$^{-1}$\,Mpc$^{-1}$~\cite{Breuval:2024lsv}, raising the discrepancy to more than $6.7\sigma$ when compared with the new SPT+ACT CMB measurement. This persistent mismatch is robust across different teams, calibrators, and Hubble-flow objects on the distance-ladder side~\cite{Freedman:2020dne,Birrer:2020tax,Anderson:2023aga,Scolnic:2023mrv,Jones:2022mvo,Anand:2021sum,Freedman:2021ahq,Uddin:2023iob,Huang:2023frr,Li:2024yoe,Pesce:2020xfe,Kourkchi:2020iyz,Schombert:2020pxm,Blakeslee:2021rqi,deJaeger:2022lit,Murakami:2023xuy,Breuval:2024lsv,Freedman:2024eph,Riess:2024vfa,Vogl:2024bum,Scolnic:2024hbh,Said:2024pwm,Boubel:2024cqw,Scolnic:2024oth,Hogas:2024qlt,Li:2025ife,Jensen:2025aai,Riess:2025chq,Newman:2025gwg,Stiskalek:2025ktq}, as well as among independent analyses probing different scales of the CMB temperature and polarization power spectra~\cite{Planck:2018vyg,Planck:2018nkj,ACT:2020gnv,ACT:2025fju,SPT-3G:2025bzu}. It has motivated a wide range of theoretical extensions of the standard cosmological model, targeting both early- and late-time physics, referring respectively to modifications of the expansion history before and after recombination (see, for example,~\cite{Murgia:2016ccp,Pourtsidou:2016ico,Nunes:2016dlj,Kumar:2016zpg,Kumar:2017dnp,DiValentino:2017iww,Yang:2018uae,DiValentino:2019ffd,Yang:2020uga,Lucca:2020zjb,DiValentino:2020leo,Kumar:2021eev,Nunes:2021zzi,Gariazzo:2021qtg,Bernui:2023byc,Mishra:2023ueo,vanderWesthuizen:2023hcl,Zhai:2023yny,Liu:2023kce,Hoerning:2023hks,Pan:2023mie,Castello:2023zjr,Forconi:2023hsj,Yao:2023jau,Garcia-Arroyo:2024tqq,Benisty:2024lmj,Silva:2024ift,Giare:2024ytc,Bagherian:2024obh,Sabogal:2025mkp,DiValentino:2016hlg,DiValentino:2017rcr,Dutta:2018vmq,vonMarttens:2019ixw,DiValentino:2020naf,Yang:2021flj,DiValentino:2021rjj,Heisenberg:2022lob,Giare:2023xoc,Adil:2023exv,Lapi:2023plb,Krolewski:2024jwj,Bousis:2024rnb,Tang:2024gtq,Manoharan:2024thb,Poulin:2018cxd,Smith:2019ihp,Niedermann:2019olb,Krishnan:2020obg,Schoneberg:2021qvd,Ye:2021iwa,Poulin:2021bjr,Niedermann:2021vgd,deSouza:2023sqp,Poulin:2023lkg,Cruz:2023lmn,Niedermann:2023ssr,Vagnozzi:2023nrq,Efstathiou:2023fbn,Cervantes-Cota:2023wet,Garny:2024ums,Giare:2024akf,Giare:2024syw,Poulin:2024ken,Pedrotti:2024kpn,Kochappan:2024jyf,Greene:2023cro,Greene:2024qis,Baryakhtar:2024rky,Lynch:2024hzh,Lee:2022gzh,Seto:2024cgo,Poulin:2023lkg,DiValentino:2019exe,Alestas:2021luu,Ruchika:2023ugh,Frion:2023xwq,Ruchika:2024ymt,Visinelli:2019qqu,Ye:2020btb,Calderon:2020hoc,Sen:2021wld,DiGennaro:2022ykp,Ong:2022wrs,Akarsu:2019hmw,Akarsu:2021fol,Akarsu:2022typ,Akarsu:2023mfb,Anchordoqui:2023woo,Adil:2023exv,Akarsu:2024qsi,Halder:2024uao,Anchordoqui:2024gfa,Gomez-Valent:2023uof,Akarsu:2024eoo,Yadav:2024duq,DiValentino:2020vnx,Silva:2025hxw,Paraskevas:2024ytz,Scherer:2025esj,Gomez-Valent:2024tdb,Akarsu:2025gwi,Soriano:2025gxd,Bouhmadi-Lopez:2025ggl,Pan:2019hac,Yang:2020zuk,Yang:2021eud,Jiang:2024xnu,Gomez-Valent:2024ejh,Specogna:2025guo,Ozulker:2025ehg,Lee:2025pzo,Hart:2017ndk,Hart:2019dxi,Knox:2019rjx,Sekiguchi:2020teg,Schoneberg:2024ynd,Toda:2024ncp}).

Although modifications to early-time physics currently appear to be the most promising avenue for addressing the $H_0$ tension, an additional and independent discrepancy within the $\Lambda$CDM framework has emerged. This tension, unrelated to the $H_0$ problem, became apparent with the first data release of the Dark Energy Spectroscopic Instrument (DESI)~\cite{DESI:2024mwx,DESI:2025zgx} and reignited interest in the possibility of late-time dynamical dark energy~\cite{DESI:2024mwx,DESI:2025zgx,Cortes:2024lgw,Shlivko:2024llw,Luongo:2024fww,Yin:2024hba,Gialamas:2024lyw,Dinda:2024kjf,Najafi:2024qzm,Wang:2024dka,Ye:2024ywg,Tada:2024znt,Carloni:2024zpl,Chan-GyungPark:2024mlx,DESI:2024kob,Bhattacharya:2024hep,Ramadan:2024kmn,Notari:2024rti,Orchard:2024bve,Hernandez-Almada:2024ost,Pourojaghi:2024tmw,Giare:2024gpk,Reboucas:2024smm,Giare:2024ocw,Chan-GyungPark:2024brx,Menci:2024hop,Li:2024qus,Li:2024hrv,Notari:2024zmi,Gao:2024ily,Fikri:2024klc,Jiang:2024xnu,Zheng:2024qzi,Gomez-Valent:2024ejh,RoyChoudhury:2024wri,Li:2025cxn,Lewis:2024cqj,Wolf:2025jlc,Shajib:2025tpd,Giare:2025pzu,Chaussidon:2025npr,Kessler:2025kju,Pang:2025lvh,Roy:2024kni,RoyChoudhury:2025dhe,Paliathanasis:2025cuc,Scherer:2025esj,Giare:2024oil,Liu:2025mub,Teixeira:2025czm,Santos:2025wiv,Specogna:2025guo,Sabogal:2025jbo,Hogas:2025ahb,Cheng:2025lod,Herold:2025hkb,Cheng:2025hug,Ozulker:2025ehg,Gialamas:2025pwv,SanchezLopez:2025uzw,Lee:2025pzo,Ormondroyd:2025iaf,Silva:2025twg,Ishak:2025cay,Fazzari:2025lzd,Lu:2025gki}. When DESI baryon acoustic oscillation (BAO) data were combined with CMB measurements, the results favored a deviation from a cosmological constant at the $2.6\sigma$ level, which increased to $2.5$–$3.9\sigma$ when supernova apparent magnitudes were also included~\cite{DESI:2024mwx}. In the second data release, DESI DR2, substantial effort was devoted to identifying potential systematics behind this discrepancy; however, the tension only intensified, reaching $3.1\sigma$ for the BAO+CMB combination and rising to $2.8$–$4.2\sigma$, depending on the supernova compilation used~\cite{DESI:2025zgx}.

On the early-time side, modifications of recombination and pre-recombination physics that shrink the sound horizon at last scattering can raise the inferred $H_0$~\cite{Knox:2019rjx,Arendse:2019hev}. Among these, varying electron mass ($m_e$) models have emerged as particularly effective~\cite{Hart:2019dxi,Sekiguchi:2020teg,Hart:2022agu,Chluba:2023xqj,Lee:2022gzh,Seto:2024cgo,Schoneberg:2024ynd,Toda:2024ncp,Toda:2025dzd,Toda:2025kcq,Wang:2025dzn}, since they coherently rescale all relevant CMB distance scales ($r_{\rm s}$, $r_{\rm D}$, $r_{\rm eq}$), preserving their ratios and thereby maintaining an excellent fit to the acoustic peak pattern while allowing for a smaller sound horizon that scales the distance measurements from BAO and the distance to the last-scattering surface simultaneously, yielding a higher $H_0$ value in better agreement with the local distance measurements.

On the late-time side, dynamical dark energy has been extensively studied as an alternative to a cosmological constant. The Chevallier–Polarski–Linder (CPL) parametrisation~\cite{Chevallier:2000qy,Linder:2002et} provides a minimal two-parameter extension to $\Lambda$CDM by replacing the cosmological constant with a barotropic perfect fluid whose dynamical equation-of-state (EoS) parameter is allowed to deviate from $w=-1$, which is associated with $\Lambda$. It is a rich parametrisation that can capture a wide variety of dynamical dark energy phenomenology despite its simplicity. The EoS parameter of CPL, being a linear function of the scale factor in the metric, naturally accommodates EoS evolutions that cross the phantom divide line (PDL), $w = -1$. Interestingly, the region of the parameter space where the PDL crossing occurs is the one preferred by data combinations including the DESI BAO measurements. While this feature is commonly encountered in the literature when dark energy is treated as an effective term encompassing all possible deviations from $\Lambda$CDM (see, e.g., Ref.~\cite{Ye:2024ywg}), it cannot be realized by a minimally coupled scalar field~\cite{Cai:2009zp}. This limitation has motivated discussions on whether a dynamical dark energy model evolving similarly to CPL but without crossing the PDL, such as the thawing or freezing field scenarios~\cite{Caldwell:2005tm,Scherrer:2007pu}, could perform comparably well or even better, and whether the apparent preference for PDL crossing is merely an artifact of the CPL form itself~\cite{Shlivko:2024llw,Wolf:2024eph,Wolf:2024stt,Wolf:2025jed,Notari:2024rti,DESI:2025fii,Keeley:2025rlg,Roy:2025cxk,Chen:2025ywv,Guedezounme:2025wav,Ozulker:2025ehg}. In particular, Ref.~\cite{Ozulker:2025ehg} introduced two modified versions of the CPL parametrisation that remove the PDL-crossing behaviour, namely CPL${}_{<a_{\rm c}}$ (capturing freezing fields) and CPL${}_{>a_{\rm c}}$ (capturing thawing fields), and found that the PDL-crossing feature is genuinely favored by current data, a result that we further investigate in this work by including these models in our analyses.

Most existing analyses of the available cosmological data have examined early- and late-time mechanisms in isolation; yet the absence of a convincing resolution to the cosmological tensions has led to approaches where deviations from the standard model at early and late times are combined~\cite{Allali:2021azp,Anchordoqui:2021gji,Khosravi:2021csn,Clark:2021hlo,Wang:2022jpo,Anchordoqui:2022gmw,Reeves:2022aoi,Yao:2023qve,daCosta:2023mow,Wang:2024dka,Toda:2024ncp}. The two can interact non-trivially: the same degeneracies that make them promising separately can either lead to improved consistency across datasets or, conversely, cancel out when combined. In fact, Ref.~\cite{Toda:2024ncp} studied the varying-$m_e$ model in combination with $\Lambda_{\rm s}$CDM~\cite{Akarsu:2019hmw,Akarsu:2021fol,Akarsu:2022typ,Akarsu:2023mfb} (an effectively late-time modification with a sign-switching cosmological constant) and found that the models do not cooperate, and their combination can perform worse than the individual models. This motivates a systematic study of whether the promising early-time extension of a varying $m_e$ can be made more effective when combined with a smoothly evolving dark energy.

In the present work, our main objective is to analyze the non-crossing CPL variants of Ref.~\cite{Ozulker:2025ehg} alongside the original model, with and without the presence of a varying $m_e$, to investigate whether the preference for PDL crossing persists and to determine whether any of these combined models offer improved amelioration of the $H_0$ tension—particularly beyond what the varying-$m_e$ extension can achieve on its own. Thus, we present the latest comprehensive constraints on $\Lambda$CDM and $\Lambda$CDM + varying-$m_e$ using \textit{Planck} 2018~\cite{Planck:2018vyg}, ACT DR6 lensing~\cite{ACT:2023kun} and SPT-3G data~\cite{SPT-3G:2023flv}, DESI DR2~\cite{DESI:2025hip} or SDSS BAO~\cite{Alam:2016hwk}, and Pantheon+ supernovae~\cite{Brout:2022vxf}. We then extend the analysis to CPL and non-crossing CPL variants~\cite{Chevallier:2000qy,Linder:2002et,Caldwell:2005tm}, both with and without varying-$m_e$.

The remainder of this paper is organised as follows. In \cref{section models}, we review the two classes of extensions we investigate: varying-$m_e$ as an early-time mechanism, and CPL (together with its non-crossing variants) as late-time parametrisations of dynamical dark energy. We outline their theoretical motivation and observational challenges. In \cref{section methodology}, we describe the implementation of these models in \texttt{CLASS}, the MCMC methodology, and the dataset combinations employed, including \textit{Planck}, ACT, SPT, DESI DR2/SDSS BAO, and Pantheon+. Our main results are presented in \cref{results}, where we constrain $\Lambda$CDM + varying-$m_e$, CPL + varying-$m_e$, and non-crossing CPL + varying-$m_e$ against the chosen datasets, quantify the impact on $H_0$ and $\Omega_m$, and highlight improvements in fit quality relative to $\Lambda$CDM. We also assess the robustness of the preferred $m_e/m_{e,0}$ shift and the resilience of the phantom-divide crossing. Finally, \cref{Conclusions} summarises our findings.

\section{Models}\label{section models}

We consider two classes of extensions to the $\Lambda$CDM framework. The first is an early-time modification, in which the electron mass is allowed to differ at early times~\cite{Hart:2019dxi,Lee:2022gzh,Seto:2024cgo}, thereby altering the physics of recombination. The second class consists of late-time dark energy extensions, namely the CPL parametrisation~\cite{Chevallier:2000qy,Linder:2002et} and its two non-crossing variants~\cite{Ozulker:2025ehg}, which enable us to isolate the role of the phantom-divide-line (PDL) crossing behaviour. We analyse these extensions both individually and in combination, allowing us to assess whether late-time dynamics can provide further improvement on the $H_0$ tension beyond the varying-$m_e$ model alone and whether the observational preference for PDL crossing persists in the presence of early-time modifications.

\subsection{Early-time modification: Varying electron mass}\label{Varying me intro section}

Early-time modifications to the $\Lambda$CDM model have been extensively investigated as potential resolutions to the Hubble tension~\citep{Poulin:2018cxd, Smith:2019ihp, Niedermann:2019olb, Krishnan:2020obg, Schoneberg:2021qvd, Ye:2021iwa, Poulin:2021bjr, Niedermann:2021vgd, deSouza:2023sqp, Poulin:2023lkg, Cruz:2023lmn, Niedermann:2023ssr, Vagnozzi:2023nrq, Efstathiou:2023fbn, Cervantes-Cota:2023wet, Garny:2024ums, Giare:2024akf, Giare:2024syw, Poulin:2024ken, Pedrotti:2024kpn, Kochappan:2024jyf, Lee:2022gzh, Greene:2023cro, Greene:2024qis, Baryakhtar:2024rky, Seto:2024cgo, Lynch:2024hzh, Toda:2024ncp, Schoneberg:2024ynd}. At the heart of these attempts lies a simple empirical fact: the angular size of the sound horizon at last scattering, $\theta_s$, is measured with exquisite precision by \textit{Planck}, which enforces a strong anticorrelation between the size of the sound horizon $r_s$ and the present-day Hubble constant $H_0$. Geometrically, this angle can be expressed as $\theta_s = r_s / D_M(a_*)$, where $r_s$ is the comoving sound horizon at recombination, $D_M(a_*)$ is the comoving angular diameter distance to last scattering, and $a_*$ is the scale factor at recombination. With $\theta_s$ effectively fixed by observations, any change in $r_s$ must be compensated by an equivalent change in $D_M(a_*)$, i.e.\ $D_M(a_*) \propto r_s$. Crucially, the distance $D_M$ is highly sensitive to $H_0$; for a spatially flat universe,
\begin{equation}
D_M(a) = c \int_a^1 \frac{\dd{a}}{a^2 H(a)} ,
\end{equation}
and, in fact, $D_M \propto H_0^{-1}$, with or without spatial curvature, if $H_0$ is treated as independent of the dimensionless density parameters. Consequently, to raise $H_0$, one must reduce the sound horizon $r_s$, making early-time physics a natural arena in which to search for viable resolutions to the tension.

However, the CMB contains far more information than just $\theta_s$, and not every modification that decreases $r_s$ preserves the remaining acoustic features. Among the various proposals, a scenario with a varying electron mass $m_e$ holds a somewhat special position due to its coherent effect on several key distance scales. Assuming that other well-measured quantities, such as the baryon-to-photon ratio, remain fixed, one finds approximately
\begin{equation}
r_s \propto r_D \propto r_{\rm eq} \propto a_*,
\end{equation}
where $r_D$ is the Silk damping scale and $r_{\rm eq}$ is the horizon size at matter–radiation equality~\cite{Hart:2017ndk,Hart:2019dxi,Knox:2019rjx,Sekiguchi:2020teg,Schoneberg:2024ynd,Toda:2024ncp}. This implies that characteristic ratios such as $r_s / r_D$ are preserved while the sound horizon is shrunk. For a comparison with other widely studied early-time modifications and their impact on these distance scales, see \cref{tab:scales_comparison}. A crucial component of this coherent rescaling lies in the dependence of the Thomson scattering cross section on the electron mass, $\sigma_T \propto m_e^{-2}$, which would not be shared by an arbitrary modification that simply alters $a_*$. As a result, the varying-$m_e$ model has proven particularly successful in extending the cosmological parameter space to better accommodate higher $H_0$, yielding a genuine upward shift in the posterior values when BAO data are also included on top of the CMB data~\cite{Hart:2017ndk,Hart:2019dxi,Schoneberg:2021qvd,Sekiguchi:2020teg,Schoneberg:2024ynd,Toda:2024ncp,Seto:2024cgo,Toda:2025kcq}.\footnote{Allowing spatial curvature to vary in addition to $m_e$ further enhances this improvement, but we do not include such extensions in this work.}

From a theoretical perspective, variations in $m_e$ can naturally arise in models where a light scalar field couples to the electron Yukawa interaction, producing a time variation in the electron mass as the scalar evolves. This is also a generic feature of conformally coupled scalar–tensor theories, where all dimensionful parameters, including $m_e$, inherit a direct dependence on the scalar field through the conformal factor~\cite{Uzan:2010pm}. Such variations of fundamental constants have been explored in a wide range of scalar–field theories. In string-inspired dilaton models~\cite{Damour:1994zq,Gasperini:2002bn}, the scalar couples universally to all gauge fields and Yukawa coupling terms, leading to time variation of both $\alpha$ and $m_e$. Similar effects arise in generic extra-dimensional scenarios~\cite{Taylor:1988nw,Smith:2025,Brax:2022vlf,Burgess:2021obw}, where the electron mass inherits time dependence through axionic and dilatonic couplings.

Runaway scalar fields and coupled quintessence scenarios~\cite{Pietroni:2005pv,Olive:2001vz,Wetterich:1987fm} can also induce varying fermion masses through couplings to the Higgs sector, with Higgs-portal interactions providing a natural particle-physics realisation~\cite{Patt:2006fw,Schabinger:2005ei,Englert:2011yb} (see also~\cite{Burrage:2023rqx} for a scalar–tensor embedding). Chameleon models~\cite{Khoury:2003aq,Brax:2004qh} and general scalar–tensor theories likewise predict conformal couplings of light scalars, in which all dimensionful parameters, such as the electron mass, rescale with the scalar field. A varying electron mass can therefore be viewed as part of this broader class of varying-constant frameworks, as discussed in~\cite{Sandvik:2001rv,Barrow:2005sv}. For phenomenological studies, one often parametrises
\begin{equation}
m_e(z) = m_{e,0}\,\big(1+\delta m_e(z)\big),
\label{eq:m_e_evolution}
\end{equation}
with $m_{e,0} \equiv 511\,\mathrm{keV}$ being the present-day measured value, and the fractional variation chosen to peak around recombination in order to have maximal impact on the sound horizon.\footnote{
As stressed by in \cite{Duff:2002vp, Duff:2001ba}, only dimensionless combinations of
constants carry invariant physical meaning. Throughout this work, our varying
$m_e$ parametrisation should be understood as shorthand for the corresponding
fractional changes in dimensionless quantities that depend on $m_e$ (such as
$m_p/m_e$ or the combinations entering recombination), while other quantities in
those ratios are held fixed. We therefore do not assign physical significance to
a variation of a dimensional constant itself.}

\begin{table*}
\centering
\begin{tabular}{lcccc}
\hline\hline
Mechanism & $r_s$ & $r_D$ & $r_{\text{eq}}$ & Main effect \\
\hline
varying-$m_e$ & $\propto a_*$ & $\propto a_*$ & $\propto a_*$ & All scales shrink together \\
EDE & $\downarrow$ & $\downarrow$ (weaker)  & $\sim$ & Primarily reduces $r_s$ \\
$\Delta N_{\rm eff}$ & $\downarrow$ & $\downarrow$ (weaker) & $\uparrow$ & More damping; higher $Y_p$ \\
varying-$\alpha$/rates & indirect & $\uparrow$ & $\sim$ & Damping tail altered \\
\hline\hline
\end{tabular}
\caption{Impact of different early-time modifications on CMB distance scales. 
A varying electron mass shifts all three scales in lockstep, unlike other models. The symbol $\sim$ indicates quantities that remain approximately unchanged.}
\label{tab:scales_comparison}
\end{table*}

The electron mass evolution is not without its own possible pitfalls and constraints. \textit{Planck} data alone actually prefer a shift in the electron mass in the \emph{opposite} direction to that required to alleviate the Hubble tension. Specifically, they show a mild preference for a \emph{lower} electron mass at early times, which increases the sound horizon $r_s$ rather than decreasing it, as would be desired. The upward shift in $H_0$ only emerges once BAO data are included: BAO data prefer lower values of $\Omega_m$ for a $\Lambda$CDM-like late-time expansion, which selects the high-$H_0$ end of the $\Omega_m$–$H_0$ degeneracy that follows the tightly constrained CMB relation $\omega_m = \Omega_m h^2 \sim \text{const}$. In this sense, the CMB alone, despite directly probing the epoch where $m_e \neq m_{e,0}$, does not drive the increase in $H_0$. Instead, the varying-$m_e$ extension enlarges the parameter space in a way that allows the BAO-preferred low-$\Omega_m$ region to be reached, which is inaccessible in baseline $\Lambda$CDM. In addition, variations in $m_e$ affect Big Bang Nucleosynthesis (BBN)~\cite{Cyburt:2015mya} and are subject to stringent astrophysical and laboratory bounds on constant variation at low redshift~\cite{Uzan:2010pm}, further tightening the viable parameter space. These considerations represent the main obstacle to varying electron-mass models fully resolving the Hubble tension.

There are also significant theoretical challenges. Varying $m_e$ requires a light scalar coupled to the Standard Model, for which the simplest, safest, and best-motivated couplings are approximately universal. If the coupling is roughly universal, for example arising from a conformal rescaling, dimensionless ratios such as $m_p/m_e$ can remain unchanged, but atomic scales like the Rydberg energy ($E_R \propto m_e \alpha^2$) and scattering coefficients like the Thomson cross section ($\sigma_T \propto \alpha^2/m_e^2$) still vary, altering the microphysics of recombination~\cite{Uzan:2010pm}. Even if the proton–electron mass ratio $m_p/m_e$ remains invariant under a universal coupling, nuclear binding energies need not necessarily rescale in the same way. As a result, the relative shifts in $m_p$, $m_e$, and the binding energies alter the ratios entering atomic spectra, leading to changes in the frequencies of absorption and emission lines as the scalar evolves, in conflict with stringent astrophysical and laboratory bounds~\cite{Uzan:2010pm}. Departures from strict universality are generic in realistic constructions (e.g., scalar dependence in gauge kinetic terms or nonuniform Yukawa couplings), in which case the QCD scale $\Lambda_{\rm QCD}$ and quark masses may not track $m_e$, leading to variations in $m_p/m_e$ and in nuclear binding energies that are tightly constrained~\cite{Sandvik:2001rv}. Moreover, a light, coupled scalar typically mediates long-range forces and induces composition-dependent accelerations of astrophysical objects, challenging weak-equivalence-principle tests unless screening mechanisms are invoked~\cite{Khoury:2003aq,Brax:2004qh}. Radiative stability is another concern: maintaining a sub-eV scalar mass with appreciable couplings to electrons is technically unnatural without protective symmetries such as scaling or shift symmetries~\cite{Patt:2006fw,Schabinger:2005ei,Englert:2011yb}.

With these observational and theoretical challenges in mind, varying-$m_e$ remains the statistically most promising early-time mechanism to alleviate the Hubble tension, standing out as the only proposal that consistently reduces the discrepancy below $3\sigma$ for the datasets considered in this paper~\cite{Sekiguchi:2020teg,Lee:2022gzh,Seto:2024cgo,Schoneberg:2024ynd}. This makes it the best-motivated early-time extension to combine with late-time modifications in the search for a more complete cosmological model. Accordingly, in our implementation, we adopt a simple prescription for the evolution of the electron mass as implemented in \texttt{CLASS}, where the difference term in \cref{eq:m_e_evolution} reads 
$\delta m_e(z) = \qty(m_e/m_{e,0} - 1)\Theta(z - 50)$, with $\Theta(z)$ being the step function, $m_{e,0}$ the present-day value of the electron mass measured to be $511\,\mathrm{keV}$, and $m_e$ a different constant value in the past ($z > 50$). Hence, the electron mass undergoes a fractional transition relative to its present-day value at a redshift of $z_{\mathrm{sw}} = 50$ and beyond, ensuring a spatially uniform constant shift in the mass across recombination.\footnote{We also note the recent interesting approach of accounting for spatial variations in the electron density in Ref.~\cite{Chluba:2025jqt}. However, this is conceptually distinct from our case, as it arises from expected clumping-induced corrections rather than from variations in fundamental constants.}

\subsection{Late-time modifications: CPL with and without phantom crossing}\label{cpl intro section}

Late-time modifications to $\Lambda$CDM have been explored extensively in attempts to reconcile the Hubble tension and improve the fit to low-redshift datasets~\cite{Murgia:2016ccp,Pourtsidou:2016ico,Nunes:2016dlj,Kumar:2016zpg,Kumar:2017dnp,DiValentino:2017iww,Yang:2018uae,DiValentino:2019ffd,Yang:2020uga,Lucca:2020zjb,DiValentino:2020leo,Kumar:2021eev,Nunes:2021zzi,Gariazzo:2021qtg,Bernui:2023byc,Mishra:2023ueo,vanderWesthuizen:2023hcl,Zhai:2023yny,Liu:2023kce,Hoerning:2023hks,Pan:2023mie,Castello:2023zjr,Forconi:2023hsj,Yao:2023jau,Garcia-Arroyo:2024tqq,Benisty:2024lmj,Silva:2024ift,Giare:2024ytc,Bagherian:2024obh,Sabogal:2025mkp,DiValentino:2016hlg,DiValentino:2017rcr,Dutta:2018vmq,vonMarttens:2019ixw,Akarsu:2019hmw,DiValentino:2020naf,Yang:2021flj,DiValentino:2021rjj,Heisenberg:2022lob,Giare:2023xoc,Adil:2023exv,Lapi:2023plb,Krolewski:2024jwj,Bousis:2024rnb,Tang:2024gtq,Manoharan:2024thb}. 
Despite their variety, most late-time scenarios face the same challenge regarding the local $H_0$ measurements. BAO measurements, if anchored to the size of the sound horizon given by the $\Lambda$CDM-like early universe, strongly restrict the ability of these models to accommodate a higher $H_0$. In fact, BAO data calibrated in this way show discordance with the supernova data if their absolute magnitude is chosen to agree with a high $H_0$ value, independent of the cosmological model; see e.g. Ref.~\cite{Poulin:2024ken}. That said, independent of the $H_0$ problem, data from established cosmological probes significantly prefer a dynamical dark energy source over a cosmological constant, with a statistical significance of $\gtrsim\!3\sigma$.

The observation that late-time modifications are hindered by BAO data when those measurements are calibrated using the $\Lambda$CDM sound horizon naturally raises a key, and arguably underexplored, question: \emph{can late-time modifications be revived when combined with early-time physics that alter $r_s$?} In particular, one may ask whether a dynamical dark energy component, when paired with an early-time mechanism that reduces the sound horizon, (i) can retain (or preferably enhance) the improvement in the $H_0$ tension provided by the early-time modification, and (ii) simultaneously offer a better fit to dataset combinations (which usually include DESI BAO) that favour late-time deviations from a cosmological constant. A related and preceding question to the latter part is whether such deviations from a cosmological constant remain preferred even in the presence of early-time modifications.

As discussed in \cref{Varying me intro section}, the early-time modification considered in this work is a sudden shift in the electron mass at redshift $z_{\rm sw} = 50$. 
For the late-time sector, we study three models: the CPL parametrisation~\cite{Chevallier:2000qy,Linder:2002et} and its two non-crossing variants introduced in Ref.~\cite{Ozulker:2025ehg}. The CPL model is a linear equation-of-state parametrisation for dark energy that can phenomenologically capture the dynamics of a wide variety of models. It is a natural benchmark for late-time extensions, particularly given that one of our central goals is to assess whether the preference for dynamical dark energy over a cosmological constant (initially observed in analyses of DESI BAO data using CPL~\cite{DESI:2024mwx,DESI:2025zgx}) persists when an early-time modification is included.

The two non-crossing variants modify the CPL form without introducing any additional parameters but restrict the evolution to avoid crossing the phantom divide line (PDL). By comparing these restricted models to the standard CPL case, one isolates the impact of PDL crossing specifically. This is an important exercise, as cosmological fits using CPL not only favour dynamical evolution but tend to prefer precisely the region of parameter space exhibiting PDL crossing. This crossing is a generic feature of CPL, but one that cannot be realised by minimally coupled scalar-field models of dark energy. In fact, the two non-crossing variants, namely $\textnormal{CPL}_{<a_{\rm c}}$ and $\textnormal{CPL}_{>a_{\rm c}}$, can be interpreted as phenomenological representations of freezing and thawing scalar-field scenarios, respectively. Their inclusion is essential to one of the main questions of this work, namely whether the preference for PDL crossing persists once early-time modifications (specifically varying-$m_e$) are introduced.

It is important to note that none of these late-time models, when considered in isolation, provide a meaningful resolution to the Hubble tension. In fact, in most cases, they exacerbate it. One might therefore question the motivation for combining such late-time models with a promising early-time mechanism like varying-$m_e$. However, Ref.~\cite{Toda:2024ncp} showed that even combining varying-$m_e$ with $\Lambda_{\rm s}$CDM, a late-time model that does alleviate the tension on its own, led to a combined outcome that performed worse than either individual model and, in terms of the inferred mean value of $H_0$, performed even worse than $\Lambda$CDM.

\begin{figure}
    \centering
    \includegraphics[width=0.45\textwidth]{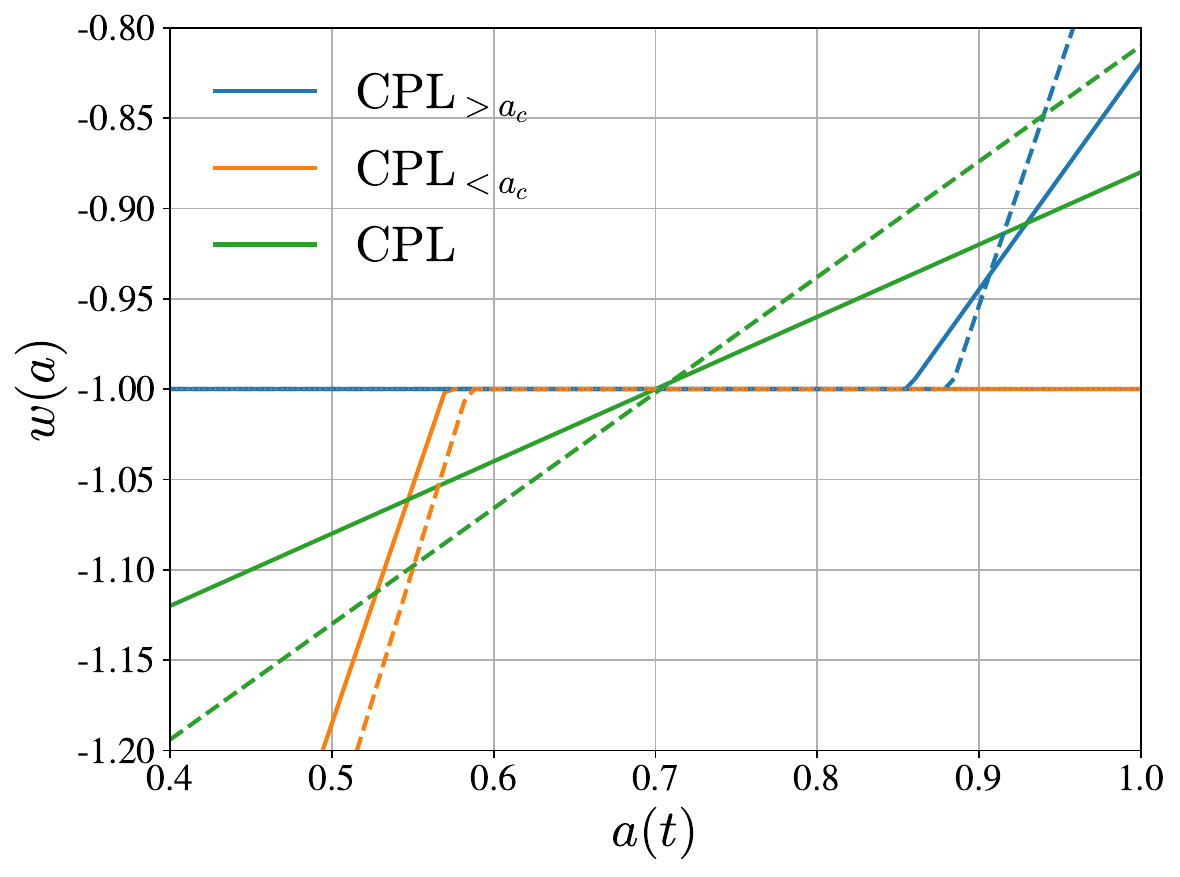}
    \caption{Evolution of the dark energy equation of state $w(a)$ for the CPL model and its non-crossing variants, shown with and without a varying electron mass. Solid lines correspond to the baseline models, while dashed lines indicate their varying-$m_e$ counterparts. Parameter values are taken from the best fits in \cref{tab:all_act_constraints}, using the CMB-A+DESI+PP dataset.}
    \label{fig:ac_best_fits}
\end{figure}

The CPL model extends the baseline $\Lambda$CDM framework by introducing two additional parameters, $w_0$ and $w_a$, replacing the cosmological constant with a barotropic perfect fluid whose equation of state varies linearly with the scale factor,
\begin{equation}
w(a) = w_0 + w_a (1 - a).
\end{equation}
This functional form generically allows a crossing of the phantom divide, $w(a) = -1$, at the scale factor
\begin{equation}\label{a_c}
a_{\rm c} \equiv 1 + \frac{1 + w_0}{w_a}.
\end{equation}
The ``non-crossing'' variants are defined by enforcing $w(a)$ to remain fixed at $-1$ whenever the vanilla CPL form would otherwise cross the PDL at $a_{\rm c}$, while behaving identically to CPL elsewhere. This leads to two possibilities, depending on whether the EoS follows CPL before or after the would-be crossing scale,
\begin{align}
\textnormal{CPL}_{>a_{\rm c}}: \quad
w(a) &=
\begin{cases}
  w_0 + w_a (1 - a), & a > a_{\rm c}, \\
  -1, & a < a_{\rm c},
\end{cases} \\[6pt]
\textnormal{CPL}_{<a_{\rm c}}: \quad
w(a) &=
\begin{cases}
  -1, & a > a_{\rm c}, \\
  w_0 + w_a (1 - a), & a < a_{\rm c}.
\end{cases}
\end{align}
Examples of $w(a)$ evolution for all three models are shown in \cref{fig:ac_best_fits}.
In the $\textnormal{CPL}_{>a_{\rm c}}$ model, the EoS is frozen at early times and evolves only after $a_{\rm c}$, whereas in $\textnormal{CPL}_{<a_{\rm c}}$ it evolves at early times but halts at the phantom divide. These constructions provide controlled departures from $\Lambda$ while explicitly forbidding phantom crossing and can be regarded as simplified limits of thawing- and freezing-type scalar-field dynamics~\cite{Caldwell:2005tm}. In this interpretation, $\textnormal{CPL}_{>a_{\rm c}}$ resembles a thawing scalar field uncoupled from matter: early-time evolution is suppressed by Hubble friction, and dynamics begin only once dark energy dominates~\cite{Notari:2024rti,Wolf:2023uno,Wolf:2024eph}. Conversely, $\textnormal{CPL}_{<a_{\rm c}}$ behaves like a freezing scenario in which the field is dynamical at early times and gradually settles toward a constant energy density after $a_{\rm c}$~\cite{Watson:2003kk,Chiba:2009gg,Dutta:2011ik}. Additionally, these parametrisations can phenomenologically describe scalar fields that are coupled to other sectors. For instance, for sufficiently flat potentials, $\textnormal{CPL}_{<a_{\rm c}}$ can mimic the behaviour of a scalar coupled to matter. In such models, the effective potential is shifted by the background density,
$V_{\rm eff}(\phi) = V(\phi) + \rho_m f(\phi)$,
so that during matter domination the field is displaced away from its vacuum configuration, producing $w(a) \neq -1$. As the matter density redshifts, the coupling term becomes subdominant and the field relaxes toward the bare potential, causing the equation of state to freeze back to $w \simeq -1$ at late times for sufficiently flat potentials. This qualitative pattern is well known in coupled quintessence and screened scalar scenarios~\cite{Wetterich:1994bg,Amendola:1999er,Khoury:2003rn,Brax:2004qh,Das:2005yj,vandeBruck:2015ida}.

While phenomenologically useful, CPL should not be mistaken for a fundamental scalar-field model. Its perturbations are treated as those of a perfect fluid, whereas in realistic scalar theories the dynamics depend on the kinetic term, potential, and couplings. Vanilla CPL cannot capture richer features such as oscillations, tracking behaviour, or sharp transitions. Its minimality is a strength but also a limitation: it cannot reproduce more complex dynamics such as thawing–freezing transitions, multi-slope potentials, or non-minimal couplings~\cite{Zhao:2017cud,DiValentino:2017zyq,Alestas:2020mvb}. The PDL-crossing behaviour inferred from data may reflect the restrictions of the parametrisation rather than genuine physics, although there is substantial support for the genuineness of the PDL crossing~\cite{Wolf:2024stt,Ozulker:2025ehg,Wolf:2025jed,Nesseris:2025lke,Gomez-Valent:2025mfl,DESI:2024aqx,Ye:2024ywg,Gonzalez-Fuentes:2025lei}. 

If one assumes a scalar-field origin for the three CPL models, a parallelism arises with the varying electron mass, as it also almost inevitably points to a scalar origin. Since $m_e = y_e v$, with $y_e$ the Yukawa coupling and $v$ the Higgs vacuum expectation value, any cosmological variation requires promoting one of these parameters to a dynamical field. This mirrors the situation in string- and extra-dimensional frameworks, where effective couplings and particle masses are controlled by moduli whose cosmological evolution induces apparent variations in the ``constants.'' In varying-$m_e$ models, the scalar couples to the electron mass and shifts atomic binding energies, altering the recombination history and the location of the CMB acoustic peaks; however, once recombination is complete, the scalar’s effect on late-time dynamics is negligible. In quintessence, by contrast, the scalar contributes directly to the dark-energy sector and can remain dynamical at late times during sustained cosmic acceleration. A natural path could be to consider a single scalar field to drive both modifications in the two different epochs. However, this does not appear to be a straightforward path: attempting to use a single scalar to both source late-time dark energy and induce electron-mass variation would generally violate stringent bounds on low-redshift variations of fundamental constants. Two distinct scalar degrees of freedom would therefore seem to be required, which is plausible in many fundamental theories, though at the cost of economy and naturalness.

Despite these caveats, CPL remains the standard benchmark\footnote{Alternative approaches include non-parametric reconstructions of $w(z)$ or model-inspired forms based on specific quintessence or k-essence potentials. These offer greater flexibility but at the cost of additional parameters and reduced interpretability.} for testing late-time dynamics and is a natural candidate to complement early-time mechanisms such as varying-$m_e$. In what follows, we therefore treat CPL and its non-crossing variants to quantify the interplay between early- and late-time physics in a model-independent way.

\section{Methodology \& Datasets}\label{section methodology}
\subsection{Model specifics}

In this paper, we consider variations of the CPL model with and without the inclusion of a varying electron mass, as well as constraints on the $\Lambda$CDM model combined with a varying-$m_e$, using the datasets listed in \cref{sec:data}. The model combinations are summarised in \cref{tab:models}. In addition to the six standard $\Lambda$CDM parameters, the varying electron mass models introduce the electron mass rescaling parameter $m_e/m_{e,0}$,\footnote{This extension is denoted simply as +$m_e$ throughout the paper.} and, in CPL-type cases, the dark energy equation-of-state parameters $(w_0, w_a)$. We impose wide, flat priors on these parameters, as shown in \cref{tab:models}. Appendix~\ref{wa priors} discusses the impact of tighter $w_a$ priors, as considered in Ref.~\cite{Schoneberg:2024ynd}.

\begin{table}
\centering
\renewcommand\arraystretch{1.12}
\begin{tabular}{l c c}
\toprule
Model & Parameters & Priors \\
\midrule
\multirow{6}{*}{$\Lambda$CDM}
  & $\Omega_b h^2$ & $[0.005, 0.1]$ \\
  & $\Omega_c h^2$ & $[0.001, 0.99]$ \\
  & $\tau_{\rm reio}$ & $[0.01, 0.8]$ \\
  & $A_s$ & $[5\times10^{-10}, 5\times10^{-9}]$ \\
  & $n_s$ & $[0.8, 1.2]$ \\
  & $\theta_s$ & $[0.5, 10]$ \\
\midrule
CPL & \makecell{$w_0$ \\ $w_a$}
    & \makecell{$[-3, 1]$ \\ $[-3, 2]$} \\
CPL$_{>a_{\rm c}}$ &  & \\
CPL$_{<a_{\rm c}}$ &  & \\
\midrule
+$m_e$ & $m_e/m_{e,0}$ & $[0.9, 1.1]$ \\
\bottomrule
\end{tabular}
\caption{Free parameters, their labels, and flat prior ranges, adopted for each model.}
\label{tab:models}
\end{table}

For the electron mass variation, we consider an instantaneous switch publicly available in \texttt{CLASS}: the electron mass is rescaled by a factor $m_e/m_{e,0}$ for all $z > z_{\rm sw}$ and restored to its standard value $m_{e,0}$ for $z < z_{\rm sw}$, thereby satisfying local constraints on the electron mass today. Our baseline choice is to implement this transition at the default redshift $z_{\rm sw} = 50$, i.e.\ after recombination and deep in the matter-dominated era, so that the primary CMB anisotropies (imprinted at $z \sim 1100$) fully experience the modified $m_e$, while late-time physics (reionization and below) does not. This choice corresponds to the treatment in~\cite{Hart:2017ndk} and is somewhat arbitrary, since the additional optical depth accumulated in this range is negligible. For the non-crossing CPL cases, we enforce the conditions $w(a) = -1$ for $a < a_{\rm c}$ or $a > a_{\rm c}$ in the $\textnormal{CPL}_{>a_{\rm c}}$ and $\textnormal{CPL}_{<a_{\rm c}}$ models, respectively, by directly modifying the equation of state of the fluid-like dark energy component in \texttt{CLASS} from the default CPL implementation. In this way, once the phantom-divide line is reached by the background dark energy equation of state, it remains fixed at $w = -1$ in the past or future, respectively.

\subsection{Datasets \& analysis}
\label{sec:data}

We explore our cosmological scenarios across several complementary datasets, spanning early- and late-time probes. In particular, we make use of the following combinations:

\noindent\textbf{\emph{Cosmic Microwave Background.}}
We consider two main combinations of CMB datasets:

\begin{itemize}
\item \textbf{Planck+ACT (CMB-A):} This dataset includes the full-mission \textit{Planck} 2018 legacy release~\cite{Planck:2018nkj,Planck:2019nip}, using the high-$\ell$ TT, TE, and EE likelihoods from \texttt{Plik}, together with the low-$\ell$ TT and EE likelihoods from \texttt{Commander} and \texttt{SimAll}, respectively. We also include the \texttt{actplanck\_baseline v1.2} likelihood, which combines CMB lensing measurements from the \textit{Planck} PR4 release and the Atacama Cosmology Telescope (ACT) DR6 data~\cite{ACT:2023kun,ACT:2023dou,Carron:2022eyg}.
    
\item \textbf{Planck+ACT+SPT (CMB-B):} This combination extends CMB-A by incorporating the new \textit{SPT-3G D1} dataset~\cite{SPT-3G:2025bzu}, which provides high-resolution measurements of both temperature and polarization anisotropies (TT, TE, EE) and an independent CMB lensing reconstruction (MUSE~\cite{SPT-3G:2025zuh}). The SPT-3G D1 analysis is based on observations of the 1500\,deg$^2$ SPT-3G Main Field from the 2019–2020 seasons, delivering the deepest ground-based CMB maps to date. The SPT-3G D1 power spectra achieve the tightest constraints on the lensed EE and TE spectra at multipoles $\ell \simeq 1800$–$4000$, surpassing ACT DR6 sensitivity at the smallest angular scales. In this work, we employ the \texttt{sptlite} version of the likelihood, which is marginalised over the nuisance parameters and provides a computationally faster implementation without loss of accuracy in the cosmological constraints.
\end{itemize}

\noindent\textbf{\emph{Baryon Acoustic Oscillations.}} We use baryon acoustic oscillation measurements from either DESI DR2 or the SDSS/eBOSS compilation. We denote these two choices by:
\begin{itemize}
  \item \textbf{DESI DR2 (DESI):} We employ the latest DESI DR2 BAO compilation~\cite{DESI:2025zpo}, which provides 16 measurements of $D_M/r_d$ and $D_H/r_d$ over $0.4 < z < 4.2$, together with a $D_V/r_d$ point at lower redshifts ($0.1 < z < 0.4$), as given in Table~IV of Ref.~\cite{DESI:2025zgx}. These measurements are derived from a mix of tracers, including ELGs, LRGs, QSOs, and Ly$\alpha$ forest data, and represent the most precise BAO distances currently available at low to intermediate redshifts.  
  \item \textbf{eBOSS/SDSS (SDSS):} As an alternative baseline, we adopt the SDSS/eBOSS compilation~\cite{eBOSS:2020yzd}, which includes both isotropic and anisotropic BAO distance and expansion-rate measurements (see Table~3 of Ref.~\cite{eBOSS:2020yzd}). These data span a range of redshifts and galaxy tracers, providing robust checks on consistency with DESI.
\end{itemize}

\noindent\textbf{\emph{Supernovae.}}
We consider the following compilation of Type~Ia supernova data:
\begin{itemize}
  \item \textbf{Pantheon+ (PP):} We use the \textit{Pantheon+} compilation~\cite{Brout:2022vxf}, which comprises 1701 light curves corresponding to 1550 unique Type~Ia supernovae, spanning the redshift range $0.001 < z < 2.26$. This dataset features improved photometric calibration and a comprehensive treatment of systematics relative to earlier releases.
\end{itemize}

Each dataset combination explored in this work is designed to probe complementary aspects of cosmology. 
The CMB anisotropies plus lensing likelihoods from \textit{Planck} precisely constrain the acoustic peaks of the CMB, anchoring the main cosmological parameters 
that govern the matter and radiation content, expansion history, sound horizon, and damping scale. 
They also provide key sensitivity to the optical depth to reionization and to inflationary parameters, 
and are included in all dataset combinations (for the list of parameters that vary, see \cref{tab:models}). 
The addition of high-resolution ground-based data from ACT and SPT extends multipole coverage 
to $\ell \sim 4000$, sharpening constraints on the damping tail and lensing amplitude—observables 
that are particularly sensitive to modifications such as a varying electron mass or dynamical dark energy. 
BAO measurements from DESI or SDSS/eBOSS directly trace the late-time expansion history and 
are crucial for breaking the geometric degeneracy between $H_0$ and the sound horizon $r_s$. 
Finally, the \textit{Pantheon+} sample provides precise distance–redshift 
measurements from Type~Ia supernovae, serving as an independent anchor on the expansion rate 
at low redshift and completing the combination of early- and late-time probes used in this work.

Parameter inference is carried out using a modified version of \texttt{CLASS}~\cite{Diego_Blas_2011} interfaced with \texttt{Cobaya}~\cite{Torrado:2020dgo} for MCMC sampling. 
We adopt the \texttt{hmcode} prescription for nonlinear corrections, with \texttt{nonlinear\_min\_k\_max=20}, and use \texttt{HyRec}~\cite{Ali_Ha_moud_2011} for recombination physics. 
To handle the generic phantom crossing in the CPL parametrisation, we employ the parametrised post-Friedmann (PPF) framework~\cite{Hu:2007pj}, which prevents the breakdown of the semiclassical fluid description of dark energy and enables a smooth, continuous transition through the phantom-divide line. 
For the variation of fundamental constants, we use the instantaneous transition prescription described above. 
Neutrino physics is modelled with $N_\mathrm{ur}=2.0328$ and $N_\mathrm{ncdm}=1$, with a total mass of $m_\nu=0.06\,\mathrm{eV}$. 
We run multiple independent MCMC chains in parallel and consider them converged once the Gelman–Rubin convergence statistic satisfies $R-1 < 0.03$ for all varied cosmological parameters. 
A conservative burn-in fraction of 30\% is discarded to ensure that spurious points in parameter space are not mistakenly retained during the initial sampling phase. 
Derived parameters such as $H_0$, $\Omega_m$, and $\sigma_8$ are computed internally from the chains. 
Posterior distributions, marginal constraints, and triangle plots are analysed using \texttt{GetDist}~\cite{Lewis:2019xzd}. 
Best-fit parameters are identified as the minimum-$\chi^2$ points explored by the converged chains, while best-fit $\Delta\chi^2$ values are obtained by comparison with the corresponding best-fit $\chi^2$ point of the converged $\Lambda$CDM chains.

To quantify the level of disagreement with the local distance-ladder measurement, we follow the overlap, or tail-probability, approach to assess tension (see, e.g.,~\cite{Raveri:2018wln} for a justification of this method). We compute the statistical tension between our posterior for $H_0$ and the SH0ES constraint ($H_0 = 73.04 \pm 1.04~\mathrm{km\,s^{-1}\,Mpc^{-1}}$~\cite{Riess:2021jrx}, treated as Gaussian), following the standard procedure for evaluating tensions involving non-Gaussian posteriors~\cite{Raveri:2021wfz}. Rather than assuming Gaussian posteriors for our chains, we make use of the full posterior distribution: for each MCMC sample $H_0^{(i)}$ with normalized weight $w_i$, we compute the cumulative distribution function (CDF) of the SH0ES Gaussian at that value. The effective two-tailed probability is then given by
\begin{equation}
    p = \sum_i w_i \, \mathrm{CDF}\!\left(H_0^{(i)}\right),
\end{equation}
where we assume that the mean $H_0$ obtained from our chains is lower than the SH0ES mean. This probability is subsequently converted into an equivalent Gaussian significance, which we report as the ``tension'' in units of~$\sigma$. This method accounts for any non-Gaussian features in the posterior and provides a robust, model-independent measure of consistency with the SH0ES determination.

\section{Results}\label{results}

\begin{table*}
\centering
\caption{Constraints at the 68\% confidence level for selected parameters, with best-fit values in parentheses for $w_0$ and $w_a$. 
Results are shown for CMB-A combinations with either DESI or SDSS, with and without PP supernovae. 
The quantity $\Delta\chi^2_{\mathrm{best}}$ is reported relative to the $\Lambda$CDM baseline using the same dataset. 
Compared to CMB-B, the CMB-A combinations favour similar values of $H_0 \simeq 68$–69~km\,s$^{-1}$\,Mpc$^{-1}$ but yield weaker $\Delta\chi^2$ improvements, indicating that CMB-A does not drive a strong preference for CPL extensions. 
Sensitivity to the BAO choice (DESI vs.\ SDSS) is most pronounced for $\textnormal{CPL}_{<a_{\rm c}}$, whereas $\textnormal{CPL}_{>a_{\rm c}}$ remains more stable, consistent with the trends shown in \cref{fig:desi_vs_sdss}.}
\label{tab:all_act_constraints}
\begin{tabular}{lc ccccc}
\toprule
Model & Dataset & $m_e/m_{e,0}$ & $H_0$(km\,s$^{-1}$\,Mpc$^{-1}$) & $w_0$ & $w_a$ & $\Delta\chi^2_{\mathrm{best}}$ \\
\midrule
$\textnormal{CPL}_{>a_{\rm c}}$ + $m_e$ & CMB-A+DESI+PP & $1.0114 \pm 0.0051$ & $68.83 \pm 0.82$ & $-0.811^{+0.11}_{-0.093}$ (-0.69) & $w_{a} < -1.12$ (-2.64) & $-5.3$ \\
 & CMB-A+SDSS+PP & $1.0081 \pm 0.0063$ & $68.40 \pm 1.00$ & $-0.85^{+0.13}_{-0.10}$ (-0.80) & $w_{a} < -1.07$ (-2.87) & $-3.1$ \\
 & CMB-A+DESI & $1.0124 \pm 0.0052$ & $67.9^{+2.7}_{-2.1}$ & $-0.73 \pm 0.28$ (-0.65) & $w_{a} < -1.00$ (-1.15) & $-4.4$ \\
\addlinespace[2pt]\cmidrule(lr){1-7}\addlinespace[2pt]
$\textnormal{CPL}_{>a_{\rm c}}$ & CMB-A+DESI+PP &  & $67.62 \pm 0.57$ & $-0.830^{+0.12}_{-0.089}$ (-0.82) & $w_{a} < -1.19$ (-1.25) & $-0.4$ \\
 & CMB-A+SDSS+PP &  & $67.29 \pm 0.54$ & $-0.89^{+0.12}_{-0.10}$ (-0.72) & $w_{a} < -1.04$ (-2.62) & $-1.6$ \\
\midrule
$\textnormal{CPL}_{<a_{\rm c}}$ + $m_e$ & CMB-A+DESI+PP & $1.0045^{+0.0081}_{-0.0073}$ & $69.15 \pm 0.80$ & $-0.32^{+0.93}_{-0.33}$ (0.21) & $w_{a} < -1.24$ (-2.91) & $-5.1$ \\
 & CMB-A+SDSS+PP & $0.9926^{+0.0080}_{-0.0096}$ & $67.60 \pm 1.10$ & $-0.18^{+0.36}_{-0.28}$ (-0.24) & $w_{a} < -1.96$ (-2.24) & $-6.9$ \\
 & CMB-A+DESI & $1.0066^{+0.0082}_{-0.0065}$ & $69.48 \pm 0.80$ & $-0.37^{+1.1}_{-0.29}$ (-0.14) & $w_{a} < -1.04$ (-1.83) & $-6.6$ \\
\addlinespace[2pt]\cmidrule(lr){1-7}\addlinespace[2pt]
$\textnormal{CPL}_{<a_{\rm c}}$ & CMB-A+DESI+PP &  & $68.74 \pm 0.39$ & $-0.15^{+0.49}_{-0.36}$ (0.13) & $w_{a} < -1.63$ (-2.63) & $-5.8$ \\
 & CMB-A+SDSS+PP &  & $68.33 \pm 0.58$ & $-0.09^{+0.48}_{-0.37}$ (0.19) & $w_{a} < -1.71$ (-2.56) & $-5.9$ \\
\midrule
$\textnormal{CPL}$ + $m_e$ & CMB-A+DESI+PP & $1.0068 \pm 0.0070$ & $68.25 \pm 0.85$ & $-0.86 \pm 0.06$ (-0.81) & $-0.44^{+0.26}_{-0.24}$ (-0.64) & $-8.5$ \\
 & CMB-A+SDSS+PP & $0.9973^{+0.0091}_{-0.010}$ & $67.40 \pm 1.10$ & $-0.856^{+0.062}_{-0.070}$ (-0.92) & $-0.65^{+0.39}_{-0.30}$ (-0.39) & $-4.3$ \\
 & CMB-A+DESI & $0.9999^{+0.0061}_{-0.0075}$ & $63.9^{+1.9}_{-2.6}$ & $-0.44^{+0.26}_{-0.21}$ (-0.51) & $-1.69^{+0.68}_{-0.82}$ (-1.45) & $-9.5$ \\
\addlinespace[2pt]\cmidrule(lr){1-7}\addlinespace[2pt]
$\textnormal{CPL}$ & CMB-A+DESI+PP &  & $67.64 \pm 0.60$ & $-0.84 \pm 0.05$ (-0.88) & $-0.60^{+0.21}_{-0.19}$ (-0.40) & $-6.9$ \\
 & CMB-A+SDSS+PP &  & $67.69 \pm 0.64$ & $-0.86 \pm 0.06$ (-0.86) & $-0.57 \pm 0.25$ (-0.56) & $-5.0$ \\
\midrule
$\Lambda\textnormal{CDM}$ + $m_e$ & CMB-A+DESI+PP & $1.0107 \pm 0.0050$ & $69.62 \pm 0.69$ &  &  & $-2.6$ \\
 & CMB-A+SDSS+PP & $1.0063 \pm 0.0057$ & $68.61 \pm 0.98$ &  &  & $-1.1$ \\
 & CMB-A+DESI & $1.0116 \pm 0.0049$ & $69.85 \pm 0.69$ &  &  & $-4.2$ \\
\addlinespace[2pt]\cmidrule(lr){1-7}\addlinespace[2pt]
$\Lambda\textnormal{CDM}$ & CMB-A+DESI+PP &  & $68.30 \pm 0.28$ &  &  & $0.0$ \\
 & CMB-A+SDSS+PP &  & $67.63 \pm 0.39$ &  &  & $0.0$ \\
 & CMB-A+DESI &  & $68.39 \pm 0.29$ &  &  & $0.0$ \\
\midrule
\bottomrule
\end{tabular}
\end{table*}

In this section, we present the results and constraints obtained from our model implementations in \texttt{CLASS}. We first examine the constraints derived using the CMB-A combination, which represents the more established and widely used dataset for this early-time probe. We combine CMB-A with DESI BAO and supernova data to break the parameter degeneracies introduced by the additional degrees of freedom in our models. To test the robustness of our findings, we also replace the DESI BAO measurements with the SDSS compilation and assess which results remain stable under this change and which exhibit sensitivity to the choice of BAO dataset. The corresponding 68\% confidence intervals for the models and dataset combinations of interest are reported in \cref{tab:all_act_constraints}. We then investigate how these results change when replacing CMB-A with the CMB-B combination, which includes the new SPT-3G D1 data. The addition of SPT provides higher angular resolution but covers a smaller sky area, and it has recently been reported to increase the Hubble tension to the $>6\sigma$ level~\cite{SPT-3G:2025bzu}. The corresponding results are presented in \cref{tab:all_spa_constraints}.

\subsection{Goodness-of-fit of the models}

Including any of the three CPL parametrisations on top of varying-$m_e$ improves the overall fit across all our runs compared to $\Lambda$CDM+$m_e$. In contrast, adding the varying-$m_e$ parameter to the CPL models produces only marginal changes in the total $\chi^2$, with shifts that can occur in either direction depending on the specific dataset combination. The main exception is the CPL${}_{>a_{\rm c}}$ model in the two runs that include SPT data (see \cref{tab:all_spa_constraints}), where the addition of the electron-mass parameter yields a significant improvement of $\Delta\chi^2 > 5$. When comparing the three CPL models among themselves, the vanilla CPL consistently outperforms the two non-crossing variants across all dataset combinations in which DESI BAO measurements are used, both with and without varying-$m_e$. This indicates a robust preference in the data for the PDL-crossing feature.

A number of CPL+$m_e$ variants achieve noticeable improvements in fit relative to the plain $\Lambda$CDM baseline, with $\Delta\chi^2_{\mathrm{best}}$ decreasing by between about 3 and 9, depending on the model. Specifically, for the CMB-A+DESI+PP combination, we find that $\textnormal{CPL}+m_e$ improves by $-8.5$, $\textnormal{CPL}_{>a_{\rm c}}+m_e$ by $-5.3$, and $\textnormal{CPL}_{<a_{\rm c}}+m_e$ by $-5.1$. When using SDSS BAO data (CMB-A+SDSS+PP), the same hierarchy appears but with smaller improvements: $\textnormal{CPL}_{<a_{\rm c}}+m_e$ gives $-6.9$, $\textnormal{CPL}+m_e$ gives $-4.3$, and $\textnormal{CPL}_{>a_{\rm c}}+m_e$ gives $-3.1$. For comparison, $\Lambda$CDM+$m_e$ achieves only $-2.6$ and $-1.1$ for the two dataset combinations, respectively. It is important to note that adding additional parameters naturally allows for some improvement in $\chi^2$ simply due to the increased number of degrees of freedom. The $\Lambda$CDM+$m_e$ model introduces one extra parameter, so an improvement of $\Delta\chi^2 \sim 2$ by itself does not indicate a preference for this model when the Akaike Information Criterion (AIC), which penalizes every extra parameter by 2, is considered. Slightly better, the CPL+$m_e$ models include three additional parameters relative to $\Lambda$CDM, so improvements larger than $\Delta\chi^2 \simeq 6$ can be considered meaningful. In this context, the observed reduction of $\chi^2$ by up to $\sim 8.5$ for the CPL+$m_e$ model indicates an improvement in the goodness of fit sufficient to surpass the Akaike criterion, despite the increased model complexity.

For the cases where the supernova data are excluded (CMB-A+DESI), the additional freedom is rewarded even more strongly, with $\textnormal{CPL}+m_e$ improving by $-9.5$, $\textnormal{CPL}_{<a_{\rm c}}+m_e$ by $-6.6$, and $\textnormal{CPL}_{>a_{\rm c}}+m_e$ by $-4.4$, while $\Lambda$CDM+$m_e$ reaches only $-4.2$. These gains indicate that the extra parameters are statistically tolerated by the data and, in some cases, rewarded with a substantially better fit. The non-crossing CPL variants, in particular, yield more stable improvements across different BAO dataset choices. However, the absolute magnitude of these $\chi^2$ reductions is still not large enough to be regarded as decisive, and, importantly, they do not correspond to an upward shift in $H_0$: the lower $\chi^2$ values are accompanied by higher $\Omega_m$ and a suppressed $H_0$.

\begin{figure}[hbtp!]
        \centering
        \includegraphics[width=0.45\textwidth]{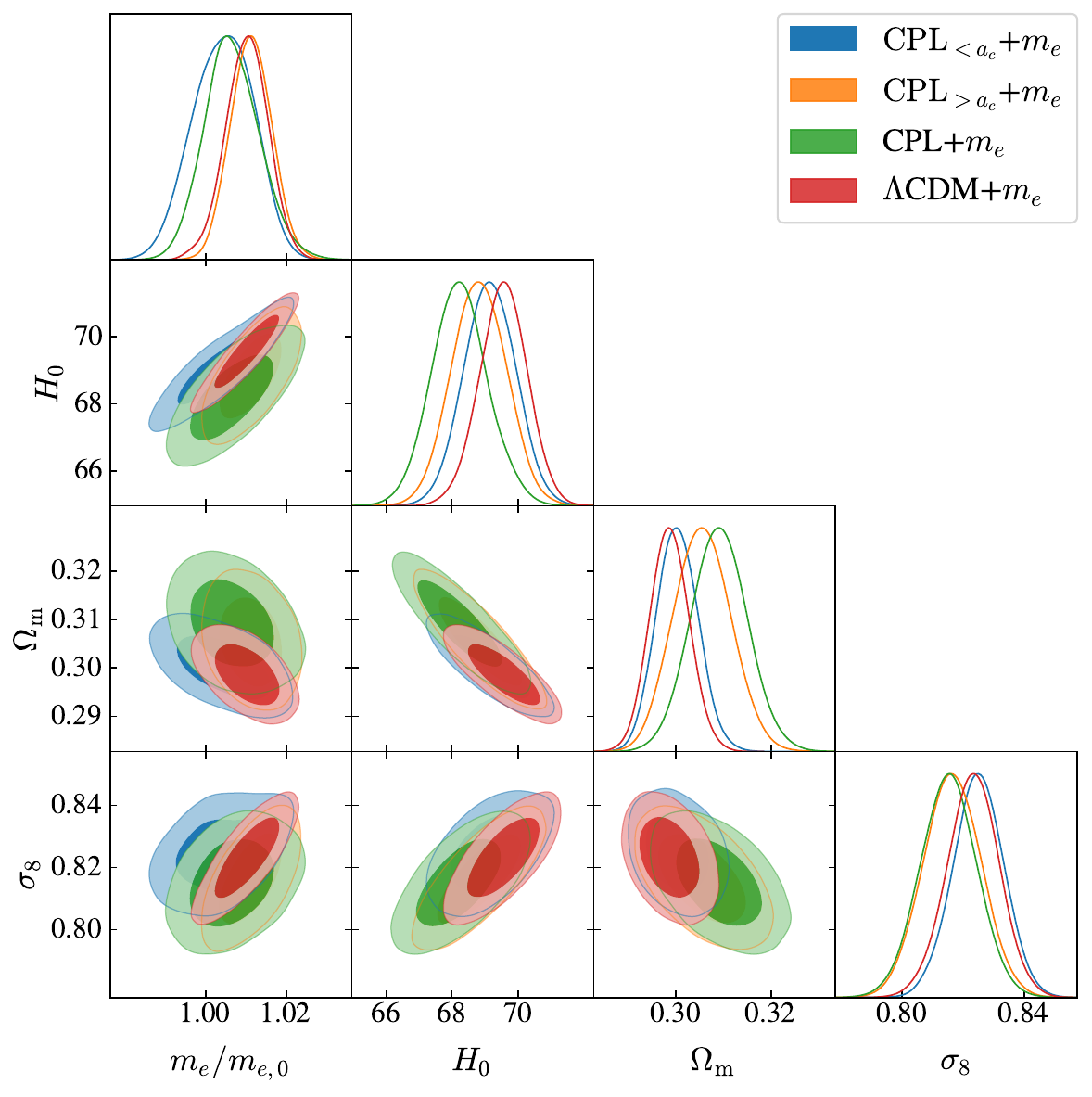}
    \caption{Triangle plot comparing parameter constraints for the four models with a varying electron mass: $\Lambda$CDM+$m_e$, CPL+$m_e$, $\textnormal{CPL}_{<a_{\rm c}}$+$m_e$, and $\textnormal{CPL}_{>a_{\rm c}}$+$m_e$. Results are based on the CMB-A+DESI+PP dataset combination, showing that all models share the same correlation directions among $m_e$, $H_0$, $\Omega_m$, and $\sigma_8$ but differ in the tightness of their constraints.}
    \label{fig:triangle CMB DESI2 PP}
\end{figure}

\subsection{Impact of the models on the $\boldsymbol{H_0}$ tension}

None of the model variants considered here provide a better resolution of the Hubble tension than the $\Lambda$CDM+$m_e$ model when constrained with the CMB-A+DESI+PP dataset combination. The most promising case remains $\Lambda$CDM+$m_e$, which yields $H_0 = 69.6 \pm 0.7$, lying at roughly $2.7\sigma$ from the SH0ES determination of $H_0 = 73.04 \pm 1.04$~\cite{Riess:2021jrx}. Among the CPL variants, the best-performing case in terms of raising $H_0$ is $\textnormal{CPL}_{<a_{\rm c}}$+$m_e$, which gives $H_0 = 69.2 \pm 0.8$, corresponding to a $3.0\sigma$ tension with SH0ES. 
As shown in \cref{tab:all_act_constraints}, the various CPL extensions with varying-$m_e$ perform only marginally worse, yielding slightly lower mean values of $H_0$ but broader posteriors that partly compensate through increased uncertainties. For instance, for the CMB-A+DESI+PP data, $\textnormal{CPL}_{>a_{\rm c}}$+$m_e$ gives $H_0 = 68.83 \pm 0.82$, $\textnormal{CPL}_{<a_{\rm c}}$+$m_e$ gives $69.15 \pm 0.80$, and $\textnormal{CPL}$+$m_e$ gives $68.25 \pm 0.85$, showing that the broader error bars somewhat compensate for the smaller means in the CPL model variants.
The direction of correlation among the parameters $m_e$, $H_0$, $\Omega_m$, and $\sigma_8$ remains the same across all models, indicating that the underlying physical relationships driving these degeneracies are unchanged. For example, while the mean $m_e$ values for $\Lambda$CDM+$m_e$ and $\textnormal{CPL}_{>a_{\rm c}}$+$m_e$ are nearly identical ($m_e = 1.0107 \pm 0.0050$ vs.\ $1.0114 \pm 0.0051$ with CMB-A+DESI+PP), the latter predicts a lower $H_0$ ($68.83 \pm 0.82$ vs.\ $69.62 \pm 0.69$), consistent with the higher preferred $\Omega_m$ seen in the triangle plot. The mean $\Omega_m$ for $\Lambda$CDM+$m_e$ is $\Omega_m = 0.2987 \pm 0.0041$, while for $\textnormal{CPL}_{>a_{\rm c}}$+$m_e$ it is $\Omega_m = 0.3056 \pm 0.0060$. In this respect, we conclude that the CPL models represent a step in the wrong direction regarding late-time modifications that might enhance the ability of a varying-$m_e$ to alleviate the Hubble tension.

\subsection{Why late-time modifications weaken the $\boldsymbol{H_0}$ alleviation from varying-$\boldsymbol{m_e}$}
\label{subsec:weakened_combined}

We should first understand how the varying-$m_e$ model can address the $H_0$ tension before investigating the effect of late-time deviations from the cosmological constant when combined. It is natural that the additional uncertainty associated with the extra free parameter $m_e/m_{e,0}$ will enlarge the uncertainty of the constraints on $H_0$ compared to $\Lambda$CDM. This behaviour is generic to any model that extends the standard cosmology but is far from adequate in resolving the tension, particularly when comprehensive datasets are taken into account that break the parameter degeneracies of the models. However, the varying-$m_e$ model also shifts the mean of the $H_0$ posterior to higher values.

It was discussed in \cref{Varying me intro section} that a larger electron mass in the early universe can shrink the sound horizon, leading to a higher $H_0$ value; however, a smaller mass is also allowed in the model, which would worsen the $H_0$ tension, and it is not immediately clear which region the data would favour. In fact, when the varying-$m_e$ model is constrained using only \textit{Planck} CMB data, the majority of the posterior volume lies in the $m_e/m_{e,0}<1$ region, pulling the $H_0$ constraints down; see, e.g., the results in Refs.~\cite{Hart:2019dxi,Baryakhtar:2024rky}. However, when the late-time data, in particular DESI BAO, are included in the dataset on top of the CMB, this situation reverses, and the $m_e/m_{e,0}>1$ region is preferred or at least contains a significantly larger portion of the posterior volume, as can be seen in \cref{tab:all_act_constraints,tab:all_spa_constraints} and also in \cref{fig:triangle CMB DESI2 PP}.

This is closely tied to the anti-correlation between $H_0$ and the dimensionless matter density parameter $\Omega_m \equiv \omega_m/h^2$, where $\omega_m$ is the physical matter density and $h$ is the reduced Hubble constant, because the impact of late-time data can be interpreted as propagating through this parameter. Ref.~\cite{Sekiguchi:2020teg} finds $m_e/m_{e,0}\propto\omega_m$ in the $\Lambda$CDM+$m_e$ model based on precisely constrained CMB observables. Moreover, they find the approximate relation $h\propto \omega_m^{3.23}$, which also yields $\Omega_m \propto (m_{e}/m_{e,0})^{-5.46}$. This last proportionality makes it clear that a low $\Omega_m$ value would lead to a high $m_e/m_{e,0}$, which in turn is able to raise $H_0$. Thus, the $\Omega_m$–$m_e/m_{e,0}$ degeneracy that opens up the parameter space considerably when only the \textit{Planck} CMB data are considered is broken by the DESI BAO data, which by itself prefer low $\Omega_m$ values compared to \textit{Planck} within a $\Lambda$CDM-like late-time cosmology (compare the results in Tab.~5 of Ref.~\cite{DESI:2025zgx} with Tab.~2 of Ref.~\cite{Planck:2018vyg}). Although Pantheon+ prefers higher $\Omega_m$ values by itself~\cite{Brout:2022vxf}, the combination of DESI BAO and Pantheon+ as a combined late-time dataset still lies on the lower side compared to the \textit{Planck} CMB results.

\begin{figure*}
  \centering
  \subfloat{%
    \includegraphics[width=0.47\textwidth]{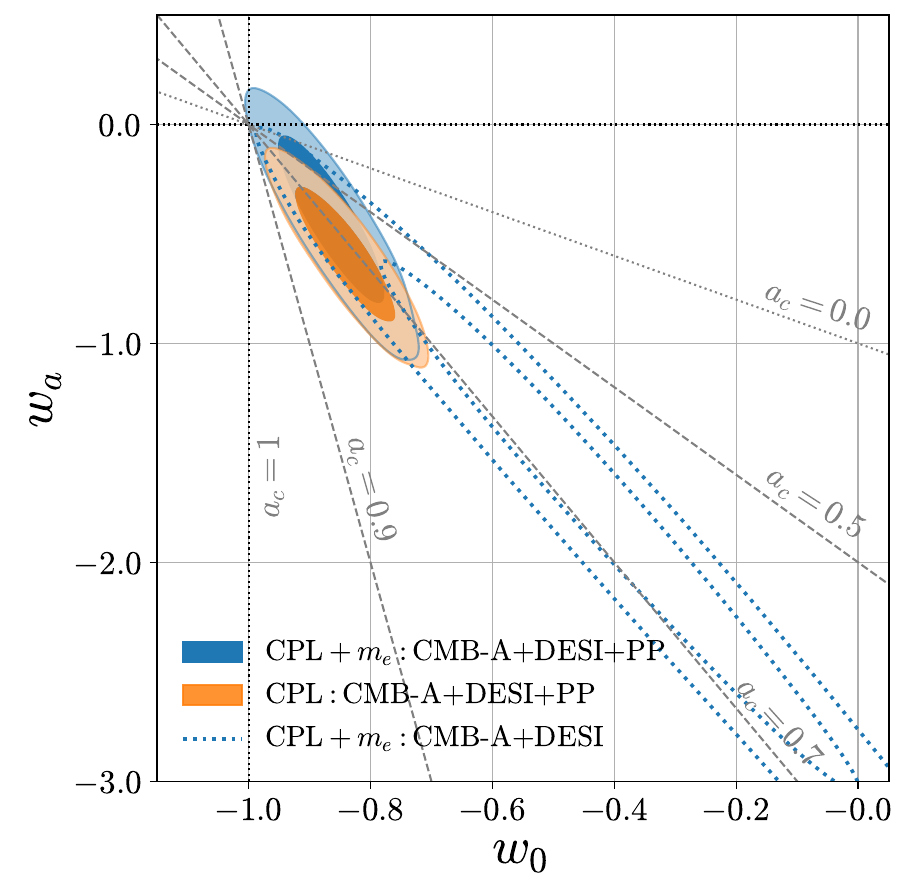}
  }
  \hfill
  \subfloat{%
    \includegraphics[width=0.47\textwidth]{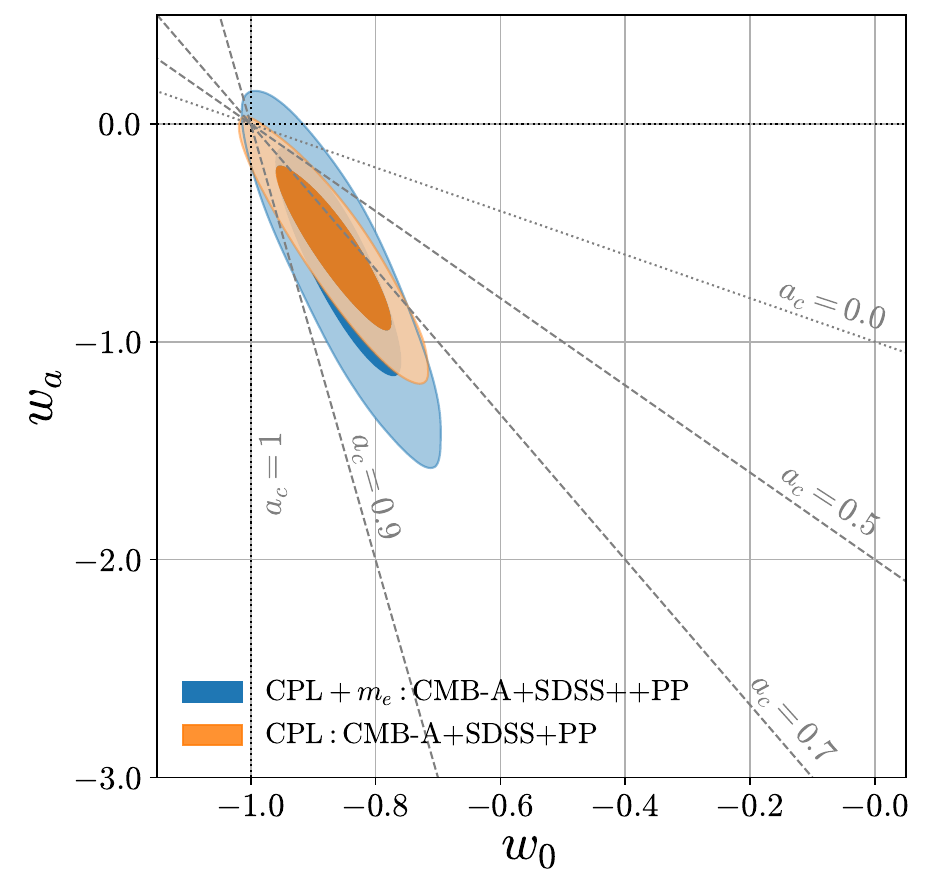}
  }
  \caption{Comparison of the posterior distributions of the CPL dark energy model with and without a varying electron mass, using CMB-A+PP combined with either DESI BAO (left) or SDSS BAO (right). Two-dimensional contours in the $w_0$–$w_a$ plane for CPL+$m_e$ (blue and dotted blue) and standard CPL (orange), with dashed lines showing example lines of constant crossing scale $a_{\rm c}$. Dotted lines illustrate $w_a = 0$ vertically and $w_0 = -1$ horizontally, with the intersection corresponding to $w(a) = -1$ at all redshifts.}
  \label{fig:cpl_me}
\end{figure*}

Allowing extra degrees of freedom at late times spoils this interplay between the datasets probing different epochs. Although the CPL parametrization yields significantly larger $\Omega_m$ values than $\Lambda$CDM when constrained by DESI BAO data alone~\cite{DESI:2025zgx}, we do not attribute the difference between the $H_0$ constraints of $\Lambda$CDM+$m_e$ and CPL+$m_e$ to this preference for higher $\Omega_m$, since the approximate scaling relation $\Omega_m \propto (m_e/m_{e,0})^{-5.46}$ is model-dependent, and its form is not straightforwardly applicable once late-time dynamics are introduced. 
Instead, we interpret our results as follows: the extra degrees of freedom in the late-time dynamical models can accommodate the requirements of the DESI BAO data within the dark energy sector itself, rather than these propagating through $\Omega_m$ to higher constraints on $m_e/m_{e,0}$. Consequently, the CMB-driven preference for lower $m_e/m_{e,0}$ values is no longer pulled upward as strongly—or, in some cases, not at all—by the complementary constraints of the late-time data. In addition, the quintessence-like behaviour of CPL for $a>a_{\rm c}$ naturally reduces $H_0$ due to a decreasing dark energy density right before the present day, and this effect is also present when the electron mass is allowed to vary. It can be seen from \cref{tab:all_act_constraints} that CPL${}_{>a_{\rm c}}$, for which this behaviour is absent, yields slightly higher $H_0$ values compared to the other two CPL parametrizations, with or without the presence of the varying-$m_e$.

Finally, we note that since the way varying-$m_e$ alleviates the $H_0$ tension is through the indirect impact of the late-time data via degeneracy breaking, combining this early modification with simple or minimal late-time dynamics is likely to lead to a weakened alleviation of the $H_0$ tension. This is also in line with the findings of Ref.~\cite{Toda:2024ncp}, where combining varying-$m_e$ with the late-time modification $\Lambda_{\rm s}$CDM~\cite{Akarsu:2019hmw,Akarsu:2021fol,Akarsu:2022typ,Akarsu:2023mfb} was counterproductive in this sense. Still, dynamics that are coupled to the matter sector or non-trivial modifications (e.g., spatial curvature that changes the functional form of the angular diameter distance and is known to work very well in combination with varying-$m_e$~\cite{Sekiguchi:2020teg}) can prove to be useful. It is also interesting to see whether a similar logic applies to the combination of late-time dynamics with other attempts to address the $H_0$ tension by modifying the sound horizon rather than the varying-$m_e$ model.

\subsection{Preference for late-time dynamics and PDL crossing in the presence of varying-$\boldsymbol{m_e}$}

In the previous subsection, we examined how introducing late-time CPL dynamics on top of the varying-$m_e$ model impacts the $H_0$ tension. Here, we take the reverse perspective and investigate how allowing $m_e$ to vary affects the conclusions drawn from the CPL model regarding late-time dynamics.

The effect on the preference for late-time dynamics is summarised in \cref{fig:cpl_me}. For the CMB-A+DESI+PP dataset analyses shown in the left panel of the figure, the addition of varying-$m_e$ reduces the preference for dynamical dark energy, as the contours widen due to increased uncertainty and also shift slightly towards the $\Lambda$CDM point corresponding to $(w_0=-1, w_a=0)$; compare the orange CPL contour with the blue CPL+$m_e$ one. The situation is similar without the inclusion of the Pantheon+ data, as can be seen by comparing the dotted blue contour of CPL+$m_e$ to the CPL contour in Fig.~11 of Ref.~\cite{DESI:2025zgx}. Still, even when varying-$m_e$ is present, $\Lambda$ is excluded at $>2\sigma$, and this exclusion is slightly stronger without the inclusion of the Pantheon+ dataset, as seen from the blue and dotted blue contours in the left panel. In comparison, when DESI BAO is replaced with SDSS BAO, the tension with $\Lambda$ is not as strong even in the CPL-only case, but curiously, the addition of varying-$m_e$ does not seem to alter the preference for dynamics significantly despite the enlarged uncertainties. The weakening of the preference for late-time dynamics when varying-$m_e$ is present can be attributed to the lack of determination of the sound horizon by the BAO measurements. The distances measured by the BAO teams are unanchored and are proportional to the inverse of the sound horizon at the drag epoch, $r_{d}^{-1}$ (which changes correspondingly with $r_s^{-1}$ for the discussions considered here), i.e., they measure $D_M/r_d$, $D_H/r_d$, and $D_V/r_d$ rather than $D_M$, $D_H$, and $D_V$. Thus, modifying the sound horizon shifts the inferred distances from the BAO data, allowing an additional path to improve the fit of the model on top of modifying the evolution of $H(z)$ with late-time dynamics. This reduces the strength of the need to deviate from $\Lambda$ at late times to accommodate the BAO data.

\begin{figure}
        \centering
        \includegraphics[width=0.47\textwidth]{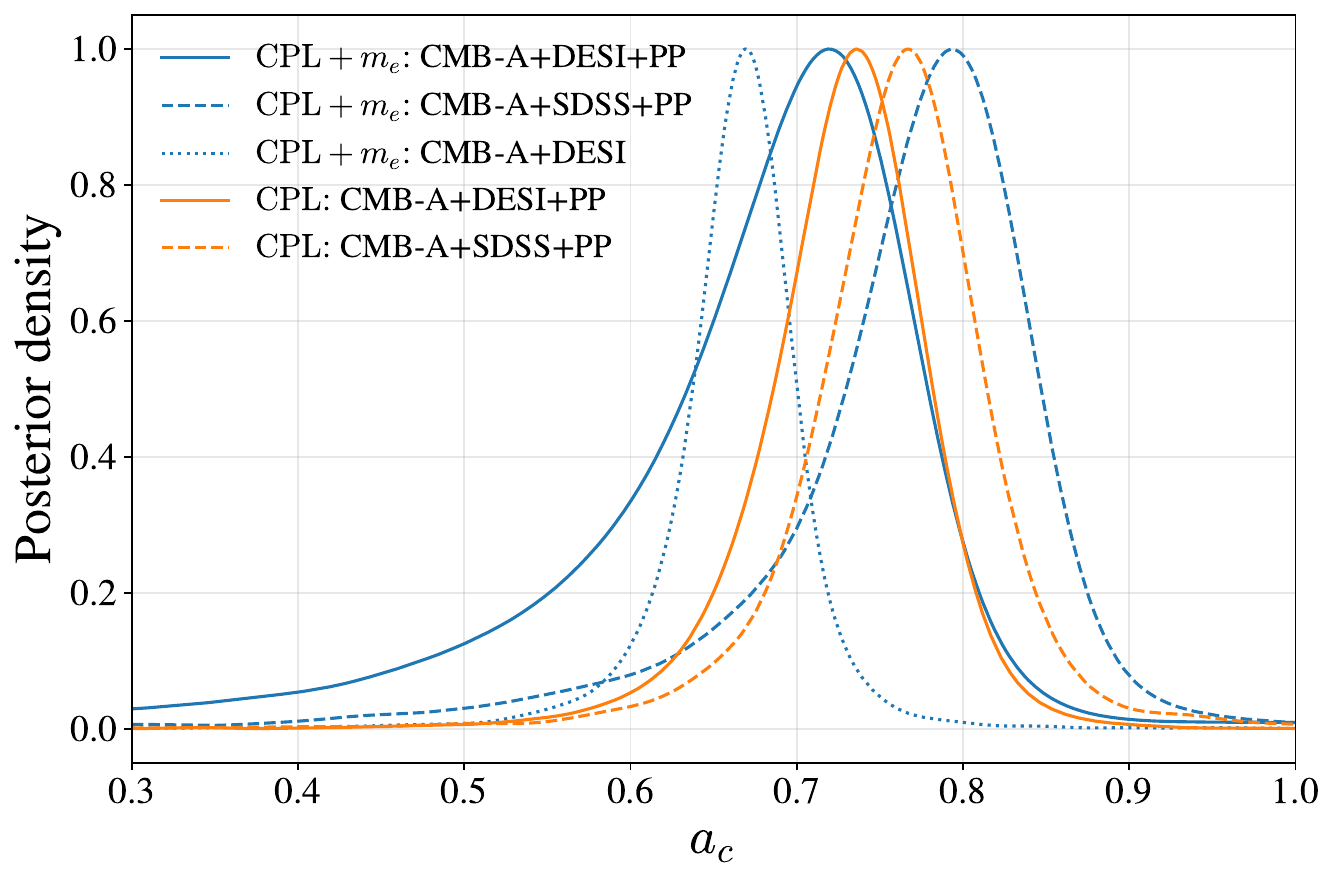}
        \caption{One-dimensional posterior distributions of the phantom-divide crossing scale factor, $a_{\rm c}$, for the CPL model, shown for different dataset combinations. Solid lines correspond to CMB-A+DESI+PP, dashed lines to CMB-A+SDSS+PP, and dotted lines to CMB-A+DESI. Orange curves represent the base CPL extension, while blue curves show CPL models including a varying electron mass.}
    \label{fig:ac_distributions}
\end{figure}

An important conclusion is the remarkable resilience of the preference for a phantom-divide crossing. The scale of phantom crossing, $a_{\rm c}$, corresponds to a degeneracy line in the $w_0$–$w_a$ plane, some examples of which are shown with dashed grey lines in \cref{fig:cpl_me}. The degeneracy directions of the CPL contours are altered only minimally by the inclusion of varying-$m_e$ in both the DESI and SDSS analyses. Despite the differing uncertainty sizes and shifts in the contours, they move and deform while closely tracking constant-$a_{\rm c}$ lines. Since the expansion history of the Universe covers the range $0<a<1$, a preference for $a_{\rm c}\in[0,1]$ indicates evidence for phantom crossing. In \cref{fig:ac_distributions}, we plot the one-dimensional marginalised posteriors of the derived parameter $a_{\rm c}$ with and without varying-$m_e$, and for both BAO datasets. Regardless of dataset choice or the inclusion of a varying electron mass, all CPL variants exhibit a clear preference for $a_{\rm c}$ within the relatively narrow range $a_{\rm c}\sim0.6$–$0.9$. Note that this preference does not directly quantify the deviation from $\Lambda$, as the cosmological constant corresponding to $w(a)=-1$ is not distinct from lines of constant $a_{\rm c}$. On the contrary, all such lines intersect at the vertex corresponding to the cosmological constant. Hence, this preference for phantom crossing implies that, given the underlying model is not $\Lambda$CDM, there is strong evidence for phantom crossing. This point is best illustrated by comparing the dotted blue CPL contour in \cref{fig:cpl_me} with the corresponding dotted blue plot in \cref{fig:ac_distributions}. Despite the contour being displaced from $\Lambda$ by slightly more than $2\sigma$, the preference for $a_{\rm c}\in[0,1]$ seen in \cref{fig:ac_distributions} is much stronger; this is because the further the contour lies from $\Lambda$, the better it tracks the $a_{\rm c}\sim0.7$ line.

Another way to quantify the preference for the PDL crossing is by comparing the goodness of fit of the vanilla CPL parametrisation with its non-crossing variants. As shown in \cref{tab:all_act_constraints}, the non-crossing variants perform worse than the vanilla CPL when the DESI BAO data are included in the analyses, with or without the varying-$m_e$. The fact that dynamical dark energy cannot perform as well when its PDL-crossing feature is removed supports the previous finding that the data favour a PDL crossing. When SDSS BAO data are considered, however, the CPL${}_{<a_{\rm c}}$ model performs slightly better. Returning to the DESI BAO case, which is more relevant for discussions of PDL crossing, we see from \cref{tab:all_spa_constraints} that the inclusion of SPT measurements in the CMB data does not change the consistent preference for the vanilla CPL over the non-crossing variants, with or without the varying-$m_e$. However, the corresponding $\Delta\chi^2$ values are smaller in this case. This weakening of the preference with the inclusion of SPT data is consistent with the findings of Ref.~\cite{Giare:2024oil}, which reported that deviations from $\Lambda$CDM towards the region of the $w_0$–$w_a$ parameter space with PDL crossing are reduced when one does not restrict the analysis to the \textit{Planck} CMB data alone.

\begin{figure}
    \centering
    \includegraphics[width=0.9\linewidth]{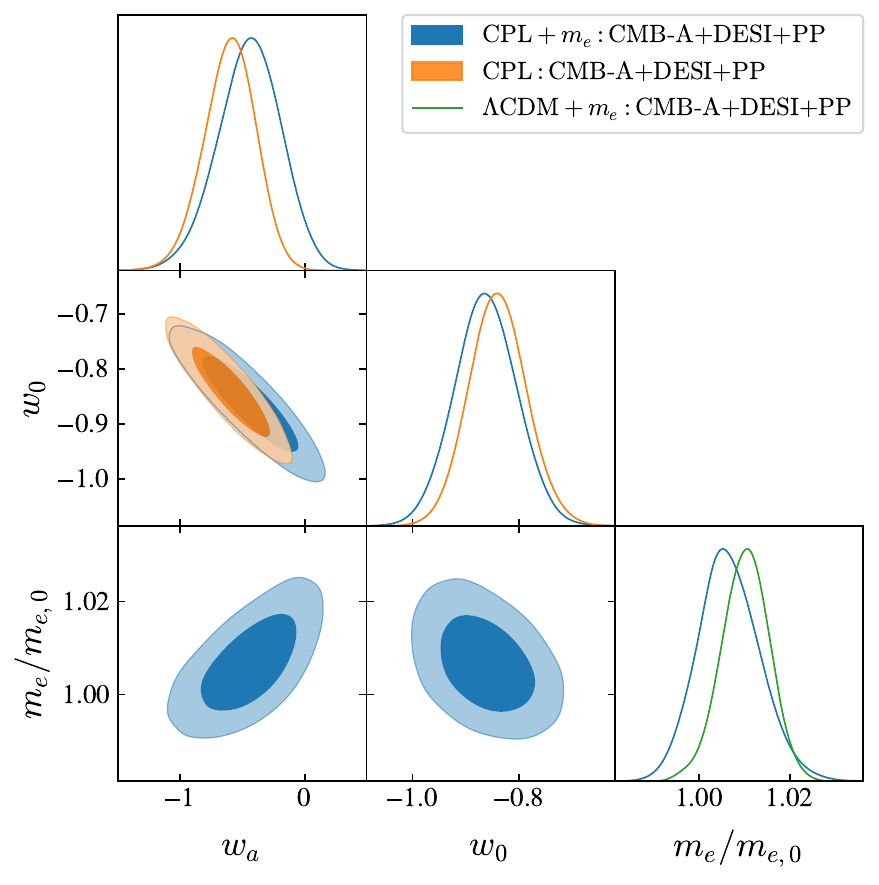}
    \caption{Triangle plot based on the CMB-A+DESI+PP data combination showing 2D contours and 1D posteriors for $\{w_0, w_a, m_e/m_{e,0}\}$, highlighting correlations in the space of extra free parameters.}
    \label{fig:comparison_all_three}
\end{figure}

Finally, we present in \cref{fig:comparison_all_three} the correlations among all three extra parameters of the CPL+$m_e$ model. In line with the above discussion, we see from the blue contours that when $m_e/m_{e,0}$ agrees with its $\Lambda$CDM value ($m_e/m_{e,0}=1$), the CPL parameters deviate from $\Lambda$, and when the CPL parameters agree with their $\Lambda$CDM values ($w_0=-1$ and $w_a=0$), the electron mass parameter deviates from unity. From a converse perspective, within the $2\sigma$ regions, the further the parameters of either the late- or early-time modifications move away from their $\Lambda$CDM values, the closer the other sector is pulled back toward $\Lambda$CDM. Altogether, the data indicate that, in the presence of a varying-$m_e$, late-time deviations from $\Lambda$ in the form of dynamical dark energy are still preferred, and evidence for PDL crossing persists across analyses. The preferred $a_{\rm c}$ values remain stable across dataset combinations, including the addition of Pantheon+, and even when recombination physics is modified by a varying electron mass.

\subsection{Different datasets}

\subsubsection*{The role of Pantheon+}

An important detail concerns the role of supernova data. As seen in \cref{tab:all_act_constraints}, Pantheon+ provides much of the constraining power for CPL-type models, effectively bringing them back in line with the other CPL variants. Including Pantheon+ in the CMB+BAO combinations both raises the inferred $H_0$ and narrows the allowed range of late-time equation-of-state parameters. Consequently, the CPL+$m_e$ models can no longer rely on inflated uncertainties to mitigate the Hubble tension. For instance, $\textnormal{CPL}+m_e$ constrained by CMB-A+DESI alone yields a low $H_0 = 63.9^{+1.9}_{-2.6}~\mathrm{km\,s^{-1}\,Mpc^{-1}}$, which increases to $68.25 \pm 0.85~\mathrm{km\,s^{-1}\,Mpc^{-1}}$ once Pantheon+ is included. Likewise, $\textnormal{CPL}_{>a_{\rm c}}+m_e$ shifts from $67.9^{+2.7}_{-2.1}~\mathrm{km\,s^{-1}\,Mpc^{-1}}$ to $68.83 \pm 0.82~\mathrm{km\,s^{-1}\,Mpc^{-1}}$ when Pantheon+ is added. In both cases, the supernova data elevate $H_0$ towards the SH0ES value, but the simultaneous tightening of the error bars maintains the tension at the $\sim 2.5$–$3\sigma$ level, comparable to the $\Lambda$CDM+$m_e$ case.

It is important to note that the Pantheon+ sample, being uncalibrated, is intrinsically agnostic to the absolute value of $H_0$. The absolute distance scale is instead set by the inverse distance ladder through the combination of CMB and BAO data. In this context, Pantheon+ does not itself drive the Hubble tension but reshapes how it appears by tightening degeneracies among late-time parameters. As seen here, adding Pantheon+ to the CMB-A+DESI dataset combination raises the inferred $H_0$ compared to CMB+DESI alone, while simultaneously narrowing the allowed parameter space. This removes the ability of CPL+$m_e$ models to rely on broad posteriors to ease the tension.  
In particular, the DESI BAO data tend to favour CPL models that are quintessence-like today ($w_0 > -1$) and phantom-like in the past ($w_a < 0$)—a behaviour that drives $H_0$ to smaller values. The inclusion of Pantheon+ counteracts this trend by pulling the evolution of $w(a)$ closer to $-1$, thereby restoring an expansion history more consistent with $\Lambda$CDM and bringing $H_0$ back towards its baseline value. By contrast, in standard $\Lambda$CDM, the $H_0$ constraint from CMB+BAO is already very tight ($\sigma_{H_0}\sim0.3$–$0.4~\mathrm{km\,s^{-1}\,Mpc^{-1}}$), so adding Pantheon+ has little effect on the posterior width. Overall, Pantheon+ strengthens the internal consistency of the fits but removes the flexibility that CPL+$m_e$ models use to weaken the tension, bringing them back to the $\sim2.5$–$3\sigma$ discrepancy already present in $\Lambda$CDM.

\subsubsection*{DESI vs SDSS}\label{DESI VD SDSS}

The role of BAO data is particularly revealing when comparing DESI and SDSS inputs. \Cref{tab:all_act_constraints} shows that the non-crossing CPL variants respond quite differently to the choice of dataset. For CMB-A+DESI+PP (DESI BAO), $\textnormal{CPL}_{<a_{\rm c}}+m_e$ sits close to $\Lambda$CDM+$m_e$, with $H_0 \simeq 69.1~\mathrm{km\,s^{-1}\,Mpc^{-1}}$ and a modest fit improvement ($\Delta\chi^2 \simeq -5$). However, switching to CMB-A+SDSS+PP (SDSS BAO) drives $H_0$ down by nearly $1.5~\mathrm{km\,s^{-1}\,Mpc^{-1}}$ and worsens the relative fit, highlighting its strong sensitivity to the BAO input. By contrast, $\textnormal{CPL}_{>a_{\rm c}}+m_e$ is comparatively stable: the shift in $H_0$ between DESI and SDSS remains within $0.5~\mathrm{km\,s^{-1}\,Mpc^{-1}}$, and $\Delta\chi^2$ changes by less than 2, indicating that the “late-evolving’’ version of the model is much less affected by which BAO compilation is used.

The plain $\textnormal{CPL}+m_e$ model also exhibits an interesting trend. With DESI, it yields the lowest $H_0$ among the variants ($68.3~\mathrm{km\,s^{-1}\,Mpc^{-1}}$), while with SDSS the value drops further to $67.4~\mathrm{km\,s^{-1}\,Mpc^{-1}}$, in both cases worsening the tension relative to $\Lambda$CDM+$m_e$. Meanwhile, the baseline $\Lambda$CDM+$m_e$ model itself is remarkably stable across DESI and SDSS, with $H_0$ shifting by less than $1~\mathrm{km\,s^{-1}\,Mpc^{-1}}$ and $\Delta\chi^2$ by less than 2. This contrast highlights that the dataset dependence arises from the impact of the BAO data on the late-time dynamics. The dependence of the CPL parameters on the BAO data is shown in \cref{fig:cpl_me}; for the non-crossing variants, we do not plot the contours for the $w_0$–$w_a$ parameters due to sharp features in the posteriors (see Fig.~2 of Ref.~\cite{Ozulker:2025ehg}), but we present the triangular plot of $m_e/m_{e,0}$ with key derived parameters, where the impact of the BAO dataset on $H_0$ and $\Omega_m$ can be clearly seen.

\begin{table*}
\centering
\caption{Constraints at the 68\% confidence level for selected parameters, with best-fit values in parentheses for $w_0$ and $w_a$. Results are shown for CMB-B combinations with and without Pantheon+ supernovae, using DESI BAO data. 
The quantity $\Delta\chi^2_{\mathrm{best}}$ is reported relative to the $\Lambda$CDM baseline using the same dataset. 
Compared to CMB-A, the CMB-B combinations yield slightly higher values of $H_0 \simeq 68$–69~km\,s$^{-1}$\,Mpc$^{-1}$ and consistently stronger $\Delta\chi^2$ improvements, indicating that the inclusion of SPT data enhances the preference for CPL-type extensions. 
Among these, $\textnormal{CPL}_{<a_{\rm c}}+m_e$ remains the most stable model across different dataset choices.}
\label{tab:all_spa_constraints}
\begin{tabular}{lc ccccc}
\toprule
Model & Dataset & $m_e/m_{e,0}$ & $H_0$(km\,s$^{-1}$\,Mpc$^{-1}$) & $w_0$ & $w_a$ & $\Delta\chi^2_{\mathrm{best}}$ \\
\midrule
$\textnormal{CPL}_{>a_{\rm c}}$ + $m_e$ & CMB-B+DESI+PP & $1.0136\pm 0.0045$ & $69.03 \pm 0.79$& $-0.815^{+0.12}_{-0.094}$ (-0.75) & $w_{a} < -1.14$ (-2.04) & $-8.1$ \\
 & CMB-B+DESI & $1.0148\pm 0.0048$ & $68.1^{+2.7}_{-2.1}$& $-0.73^{+0.32}_{-0.28}$ (-0.69) & $w_{a} < -1.06$ (-1.16) & $-8.0$ \\
\addlinespace[2pt]\cmidrule(lr){1-7}\addlinespace[2pt]
$\textnormal{CPL}_{>a_{\rm c}}$ & CMB-B+DESI+PP &  & $67.44^{+0.63}_{-0.53}$  & $-0.844^{+0.13}_{-0.098}$ (-0.72) & $w_{a} < -1.24$ (-2.78) & $-0.5$ \\
 & CMB-B+DESI &  & $68.2^{+2.3}_{-1.7}$ & $-0.95^{+0.28}_{-0.24}$ (-1.01) & $---$ (-0.05) & $-2.8$ \\
\midrule
$\textnormal{CPL}_{<a_{\rm c}}$ + $m_e$ & CMB-B+DESI+PP & $1.0064^{+0.0080}_{-0.0072}$ & $69.31 \pm 0.78$ & $-0.26^{+0.83}_{-0.35}$ (0.08) & $w_{a} < -1.34$ (-2.60) & $-8.3$ \\
 & CMB-B+DESI & $1.0080^{+0.0082}_{-0.0067}$ & $69.60 \pm 0.80$ & $-0.29^{+0.95}_{-0.35}$ (0.07) & $w_{a} < -1.20$ (-2.48) & $-8.8$ \\
\addlinespace[2pt]\cmidrule(lr){1-7}\addlinespace[2pt]
$\textnormal{CPL}_{<a_{\rm c}}$ & CMB-B+DESI+PP &  & $68.75 \pm 0.38$ & $-0.11^{+0.38}_{-0.33}$ (0.13) & $w_{a} < -1.80$ (-2.74) & $-8.2$ \\
 & CMB-B+DESI &  & $68.88 \pm 0.40$ & $-0.15^{+0.40}_{-0.31}$ (-0.68) & $w_{a} < -1.76$ (-0.86) & $-9.0$ \\
\midrule
$\textnormal{CPL}$ + $m_e$ & CMB-B+DESI+PP & $1.0084^{+0.0061}_{-0.0070}$ & $68.37 \pm 0.85$ & $-0.86 \pm 0.06$ (-0.88) & $-0.46^{+0.26}_{-0.23}$ (-0.23) & $-10.2$ \\
 & CMB-B+DESI & $1.0016^{+0.0056}_{-0.0073}$ & $64.0^{+1.8}_{-2.7}$ & $-0.44^{+0.27}_{-0.21}$ (-0.41) & $-1.70^{+0.66}_{-0.83}$ (-1.76) & $-11.7$ \\
\addlinespace[2pt]\cmidrule(lr){1-7}\addlinespace[2pt]
$\textnormal{CPL}$ & CMB-B+DESI+PP &  & $67.65 \pm 0.59$ & $-0.83 \pm 0.05$ (-0.89) & $-0.67^{+0.21}_{-0.19}$ (-0.37) & $-9.6$ \\
 & CMB-B+DESI &  & $63.5^{+1.6}_{-2.1}$ & $-0.39^{+0.22}_{-0.19}$ (-0.30) & $-1.85 \pm 0.56$ (-2.17) & $-11.9$ \\
\midrule
$\Lambda\textnormal{CDM}$ + $m_e$ & CMB-B+DESI+PP & $1.0124\pm 0.0043$ & $69.74 \pm 0.66$ &  &  & $-6.7$ \\
 & CMB-B+DESI & $1.0135\pm 0.0043$ & $70.01 \pm 0.67$ &  &  & $-7.8$ \\
\addlinespace[2pt]\cmidrule(lr){1-7}\addlinespace[2pt]
$\Lambda\textnormal{CDM}$ & CMB-B+DESI+PP &  & $68.04 \pm 0.26$ (68.25) &  &  & $0.0$ \\
 & CMB-B+DESI &  & $68.13 \pm 0.26$ (68.23) &  &  & $0.0$ \\
\midrule
\bottomrule
\end{tabular}
\end{table*}

\begin{figure}
        \centering
        \includegraphics[width=0.47\textwidth]{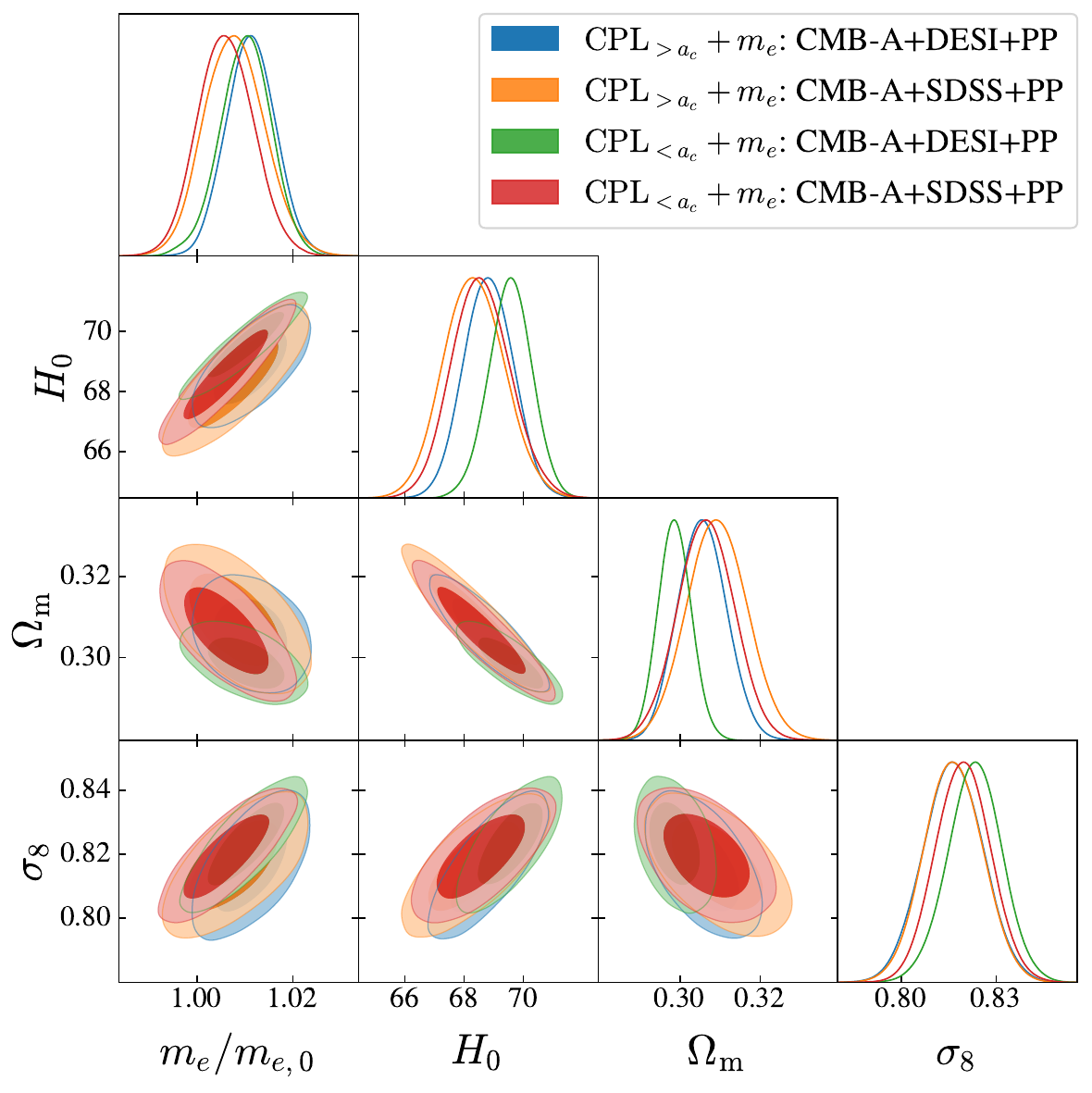}
        \caption{Triangle plot showing the 2D contour plots and 1D posterior distributions for the $\textnormal{CPL}_{>a_{\rm c}}+m_e$ and $\textnormal{CPL}_{<a_{\rm c}}+m_e$ models. Results are shown for CMB-A+PP combined with DESI (green/red) and SDSS (grey/blue) BAO data, highlighting how the early- and late-crossing CPL variants respond differently to the choice of BAO dataset.}
    \label{fig:desi_vs_sdss}
\end{figure}

\subsubsection*{Results with SPT data}

The inclusion of SPT data causes the $\chi^2$ gap between the $\Lambda$CDM and extended models to further widen, indicating stronger deviations from $\Lambda$CDM. However, as discussed in \cref{subsec:weakened_combined}, the $\chi^2$ improvements over $\Lambda$CDM+$m_e$ obtained by including late-time dynamics are milder compared to the CMB-A case, indicating a reduced preference for late-time dynamics in the presence of this dataset, in line with the findings of Ref.~\cite{Giare:2024oil}.

Without Pantheon+ supernovae, both CPL and CPL+$m_e$ drive $H_0$ to very low values, around $63$–$64~\mathrm{km\,s^{-1}\,Mpc^{-1}}$, with broad posteriors and large fit improvements ($\Delta\chi^2_{\mathrm{best}}\simeq-12$). By contrast, $\textnormal{CPL}_{<a_{\rm c}}+m_e$ stabilises $H_0$ near $69.6\pm0.8~\mathrm{km\,s^{-1}\,Mpc^{-1}}$ with CMB-B+DESI, yielding $\Delta\chi^2\simeq-8.8$ and only a $2.6\sigma$ tension with SH0ES. This combination of a higher $H_0$ and a substantially improved fit is particularly noteworthy. For comparison, $\Lambda$CDM+$m_e$ also pushes $H_0$ upward to $70.0\pm0.7~\mathrm{km\,s^{-1}\,Mpc^{-1}}$ but achieves only $\Delta\chi^2\simeq-7.8$, i.e.\ a slightly worse fit than $\textnormal{CPL}_{<a_{\rm c}}+m_e$.

Including Pantheon+ supernovae sharpens this hierarchy further. With CMB-B+DESI+PP, $\textnormal{CPL}_{<a_{\rm c}}+m_e$ gives $H_0 = 69.3\pm0.8~\mathrm{km\,s^{-1}\,Mpc^{-1}}$ ($2.9\sigma$ from SH0ES) and $\Delta\chi^2\simeq-8.3$, while $\textnormal{CPL}_{>a_{\rm c}}+m_e$ yields $H_0 = 69.0\pm0.8~\mathrm{km\,s^{-1}\,Mpc^{-1}}$ ($3.1\sigma$) and $\Delta\chi^2\simeq-8.1$. The plain CPL model also achieves strong fit improvements ($\Delta\chi^2\simeq-10$) but at the cost of driving $H_0$ much lower and pushing $w_a$ to very negative values. Interestingly, the $\textnormal{CPL}_{>a_{\rm c}}$ model benefits most from allowing $m_e$ to vary, with its fit improvement rising from near baseline to competitive with plain CPL, suggesting that modified recombination physics can partially substitute for a late-time phantom crossing.

\section{Conclusions}\label{Conclusions}

In this work, we have investigated the interplay between early- and late-time extensions to $\Lambda$CDM by combining a varying electron mass with dynamical dark-energy parametrizations. Our primary goals were to examine whether the inclusion of late-time dynamics (incorporated through the CPL model~\cite{Chevallier:2000qy,Linder:2002et} and its non-crossing variants~\cite{Ozulker:2025ehg}) affects the ability of varying-$m_e$ to alleviate the $H_0$ tension, and to test whether the preference for deviations from $\Lambda$ and the phantom-divide line (PDL) crossing observed in dynamical dark energy analyses persists once early-time modifications are introduced. To this end, we presented comprehensive constraints on $\Lambda$CDM and its extensions using \textit{Planck} 2018, ACT DR6 lensing, and SPT-3G CMB data, in combination with DESI DR2 or SDSS BAO and the Pantheon+ supernovae compilation.

Across all dataset combinations analysed, $\Lambda$CDM+$m_e$ yields the largest upward shift in $H_0$ (for example, $H_0 \simeq 69.6 \pm 0.7~\mathrm{km\,s^{-1}\,Mpc^{-1}}$ for CMB-A+DESI+PP), reducing the tension with SH0ES to about $2.7\sigma$. The CPL parametrisation exacerbates the $H_0$ tension but provides a better fit to the data compared to varying-$m_e$, and it consistently favours a late-time crossing of the phantom divide, with $a_{\rm c} \simeq 0.6$–$0.9$. This behaviour remains robust when $m_e$ is also allowed to vary: the crossing persists at a similar epoch and with comparable statistical weight, although the overall evidence for dynamical dark energy is weakened. The non-crossing variants, CPL$_{>a_{\rm c}}$ capturing the phenomenology of thawing scalar field scenarios, and CPL$_{<a_{\rm c}}$ capturing freezing scalar field scenarios, consistently perform worse in their fit to the data compared to the vanilla CPL with phantom crossing when DESI BAO is included in the dataset, whether the electron mass is allowed to vary or not, further supporting evidence for PDL crossing. Results obtained with the CMB-B combination (including SPT-3G D1) follow the same qualitative trends, albeit with weaker preference for dynamical dark energy.

Taken together, our results show that the early-time calibration provided by a percent-level increase in $m_e$ during recombination is chiefly responsible for lifting $H_0$. The varying-$m_e$ mechanism alleviates the Hubble tension mainly through the indirect impact of late-time data, which break degeneracies among $m_e$, $\Omega_m$, and $H_0$. When additional degrees of freedom are introduced at late times, as in the CPL extensions, this interplay changes: the dark-energy parameters absorb part of the role previously played by $m_e$, allowing the distance constraints to be satisfied within the dark-energy sector rather than low $\Omega_m$ values that in turn increase $m_e/m_{e,0}$ by degeneracy breaking in a $\Lambda$CDM-like cosmology. As a result, the inclusion of late-time dynamics can suppress the upward shift in $m_e/m_{e,0}$, leading to $H_0$ values that are typically lower than $\Lambda$CDM+$m_e$. This interpretation is consistent with Ref.~\cite{Toda:2024ncp}, which found that combining varying-$m_e$ with late-time modifications such as $\Lambda_{\rm s}$CDM~\cite{Akarsu:2019hmw,Akarsu:2021fol,Akarsu:2022typ,Akarsu:2023mfb} can be counterproductive. Further, in vanilla CPL and CPL$_{>a_{\rm c}}$, the quintessence-like regime right before the present day has a negative effect on the $H_0$ tension.

While the CPL parametrisation offers a convenient phenomenological handle on late-time dynamics, its implementation remains inherently ad hoc~\cite{Artola:2025zzb}. The dark-energy sector is treated as a perfect fluid with fixed sound speed, and the stability of phantom-divide crossings is imposed numerically rather than derived from fundamental physics. More consistent approaches—such as scalar-field or effective-field-theory frameworks—determine both the background and perturbation evolution from the underlying dynamics and stability conditions~\cite{Hu:2013twa}, whether motivated by a Lagrangian construction or by more general EFT principles. Canonical single-field quintessence cannot cross the phantom divide~\cite{Fang:2008sn,Hu:2004kh}, and phantom models introduce ghost instabilities~\cite{Kobayashi:2019hrl,Tsujikawa:2025wca}, while multi-field or coupled scenarios (e.g.\ quintom models~\cite{Wei:2005nw}, scalar-tensor extensions~\cite{Brax:2025ahm,Smith:2025grk,Langlois:2018dxi}, or beyond-Horndeski theories~\cite{Tiwari:2024gzo}) can realise smooth CPL-like behaviour across $w=-1$. Such setups avoid perturbative pathologies but at the cost of substantially modifying both the expansion history and structure formation, reinforcing the need for physically consistent late-time models beyond CPL and motivating their joint exploration with early-time modifications such as varying-$m_e$, as undertaken in this work.

For the early-time physics, we modelled a step-like variation of the electron mass using the instantaneous varying-constants module of \texttt{CLASS}. Although this provides a practical framework, the step approximation at $z=50$ is an idealisation that introduces a formal discontinuity in $\sigma_{\rm T}\!\propto\! m_e^{-2}$. A smoother evolution of $m_e(z)$ that perhaps changes even during recombination within a fully consistent treatment would offer a more physical implementation and will be explored in future work.

In summary, in a scenario where $m_e$ is also allowed to vary, replacing the cosmological constant with late-time dynamics (in the form of CPL and its variants) can provide modest $\Delta\chi^2$ improvements by exploiting extra flexibility in the expansion history, and they exclude the cosmological constant at $\gtrsim\!2\sigma$ with a strong preference for a PDL crossing. The evidence for the PDL crossing is further supported by the underperformance of the non-crossing CPL variants. However, the addition of the late-time dynamics does not further relieve the Hubble tension; rather, it worsens it. Consequently, $\Lambda$CDM+$m_e$ remains the configuration that most effectively raises $H_0$ among those tested. Since varying-$m_e$ alleviates the $H_0$ tension primarily through degeneracy breaking with late-time data, combining it with minimal dynamical extensions tends to weaken the improvement. More complex modifications, such as couplings to matter (spatial curvature is already known to work well alongside varying-$m_e$~\cite{Sekiguchi:2020teg}), may offer better synergy. It is also intriguing whether similar behaviour would occur when late-time dynamics are paired with other mechanisms that modify the sound horizon, and whether the preference for a dynamical dark energy and phantom crossing persists in those cases.

\begin{acknowledgments}
\noindent We thank William Giarè, Daniel Kessler, Nils Schöneberg and Jesus Torrado for their insightful suggestions and extremely useful discussions. EDV is supported by a Royal Society Dorothy Hodgkin Research Fellowship. AS is supported by the W.D. Collins Scholarship. CvdB is supported in part by the Lancaster–Sheffield Consortium for Fundamental Physics under STFC grant: ST/X000621/1. 
We acknowledge the IT Services at The University of Sheffield for the provision of services for High Performance Computing. 
This article is based upon work from the COST Action CA21136 - ``Addressing observational tensions in cosmology with systematics and fundamental physics (CosmoVerse)'', supported by COST - ``European Cooperation in Science and Technology''.
\end{acknowledgments}

\bibliographystyle{apsrev4-2}
\bibliography{bibliography}

\appendix

\section{Comparison of Priors}\label{wa priors}
\begin{table*}
\centering
\begin{tabular}{lcc}
\toprule
Parameter & $\textnormal{CPL}$ $+$ $m_e$ ($-3<w_a <2$)& $\textnormal{CPL}$ $+$ $m_e$ ($-1<w_a <1$)\\
\hline
$m_e/m_{e,0}$ & $0.9976^{+0.0070}_{-0.0089}$ & $1.0052^{+0.0081}_{-0.0093}$\\
$w_{0}$ & $-0.45^{+0.34}_{-0.18}$ & $-0.89^{+0.19}_{-0.12}$\\
$w_{a}$ & $<-1.56$ & $<-0.359$\\
$H_0$(km\,s$^{-1}$\,Mpc$^{-1}$) & $64.6^{+2.0}_{-3.5}$ & $69.0^{+1.9}_{-2.7}$\\
$\Omega_\mathrm{m} h^2$ & $0.1425^{+0.0020}_{-0.0023}$ & $0.1437\pm 0.0023$\\
$\sigma_8$ & $0.793^{+0.022}_{-0.030}$ & $0.824^{+0.022}_{-0.025}$\\
$\chi^2$ & $2790.7\pm 6.4$ & $2794.4\pm 6.3$\\
\hline
\bottomrule
\end{tabular}
\caption{Comparison of parameter constraints at the 68\% confidence level from the $\textnormal{CPL}+m_e$ model using the CMB-A+DESI dataset combination (excluding ACT DR6 lensing, unlike the rest of the paper), under different prior choices for $w_a$.}
\label{tab:priors on cpl}
\end{table*}

\begin{figure*}
  \centering
  \subfloat{%
    \includegraphics[width=0.47\textwidth]{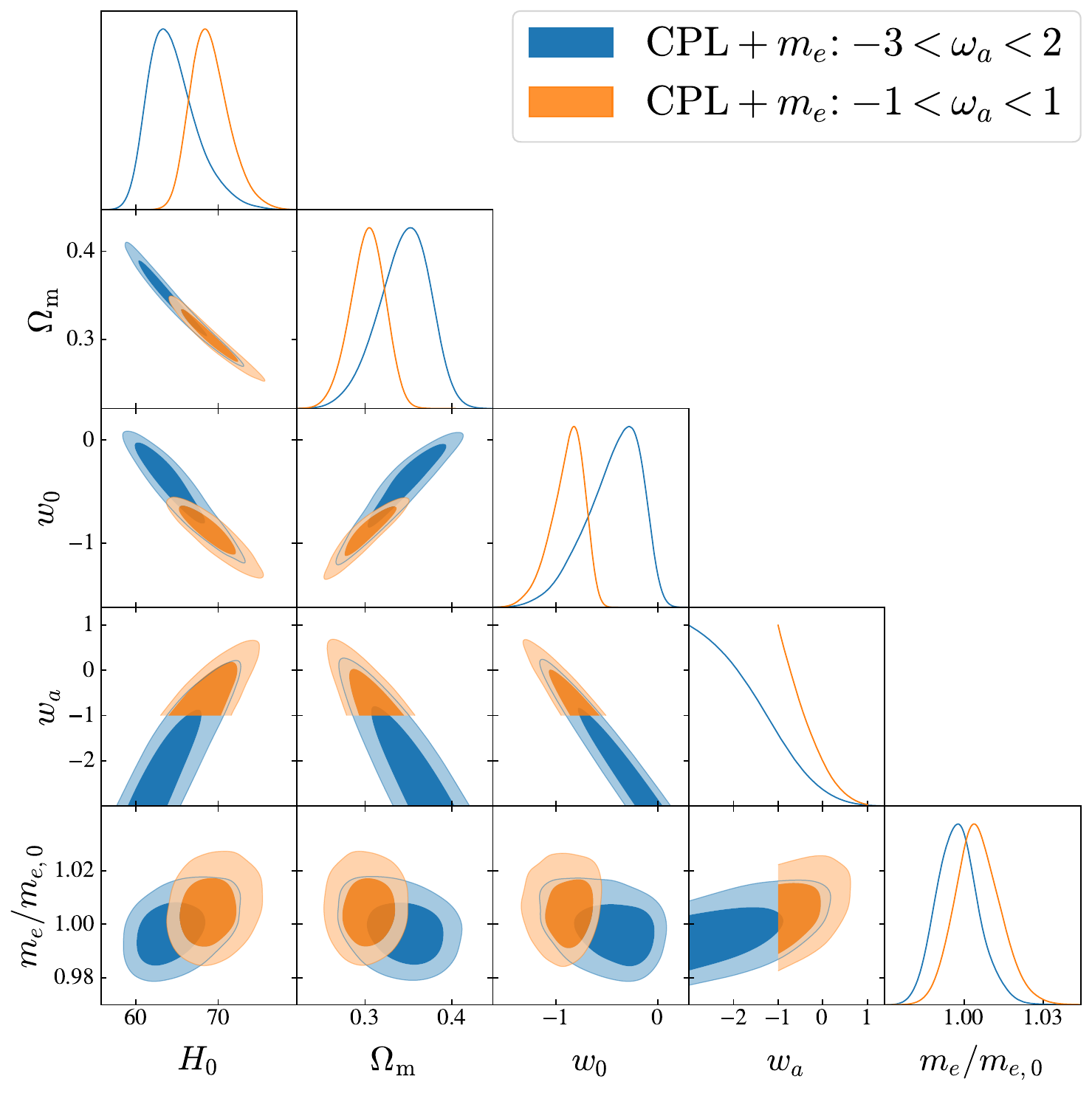}
  }
  \hfill
  \subfloat{%
    \includegraphics[width=0.47\textwidth]{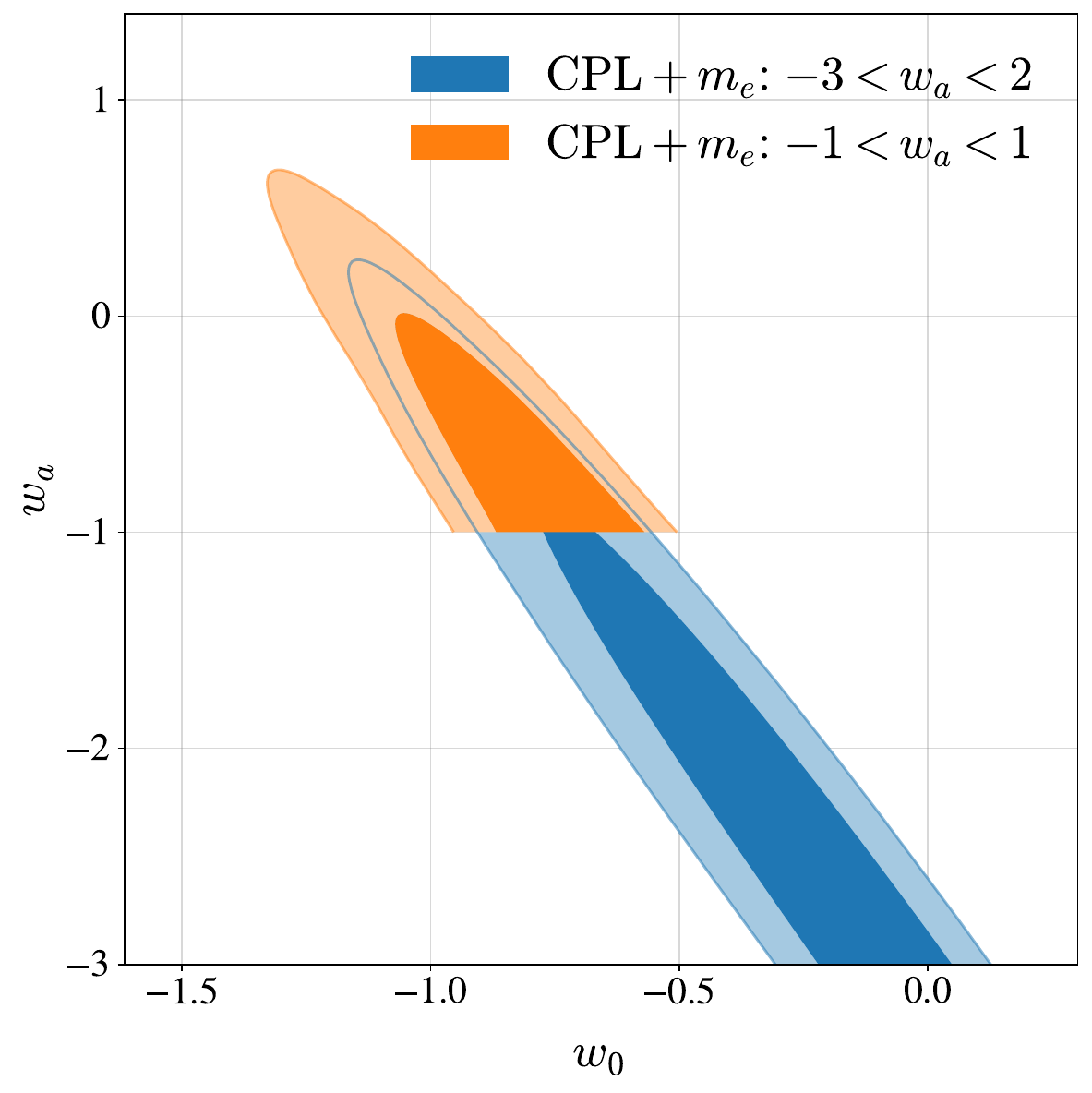}
  }
  \caption{Comparison of posterior distributions for the $\textnormal{CPL}+m_e$ model using the CMB-A+DESI dataset combination, but excluding ACT DR6 lensing. Shown are the 68\% and 95\% confidence regions for key parameters under two prior choices: wide ($-3 < w_a < 2$, blue) and narrow ($-1 < w_a < 1$, orange). The narrower prior truncates the degeneracy direction toward more negative $w_a$, thereby removing regions with lower $H_0$ and higher $\Omega_m$, and shifting the preferred parameter values.}
  \label{fig:priors_cpl_me}
\end{figure*}

Our constraints on various parameters, in particular the inferred value of $H_0$ for the CPL+$m_e$ model, differ from those obtained in~\cite{Schoneberg:2024ynd}. In this appendix, we investigate the cause of this discrepancy. For direct comparability with~\cite{Schoneberg:2024ynd}, we remove the ACT DR6 lensing data from our CMB combination and focus on the impact of the prior choice on $w_a$.

Table~\ref{tab:priors on cpl} shows the 68\% confidence intervals on the CPL+$m_e$ model when imposing a wide flat prior ($-3 < w_a < 2$) compared to the narrower prior ($-1 < w_a < 1$) adopted in~\cite{Schoneberg:2024ynd}. Allowing a wider region in the $w_a$ parameter space shifts the posterior support to more negative values. The right-hand panel of \cref{fig:priors_cpl_me} illustrates why: when constraining with CMB-A+DESI+PP datasets, $w_a$ remains unconstrained toward increasingly negative values, and imposing $w_a > -1$ artificially truncates this flat direction. This truncation affects not only $w_a$, but also propagates through its degeneracies with $H_0$ and $m_e/m_{e,0}$. More negative $w_a$ values correlate with higher $\Omega_m$ and lower $H_0$ and $m_e/m_{e,0}$, as can be seen in the elongated degeneracy directions in \cref{fig:priors_cpl_me}. Cutting the tail of this degeneracy at $w_a = -1$ removes the region with low $H_0$ corresponding to low $m_e/m_{e,0}$ and high $\Omega_m$, shifting the marginalised posteriors. It remains to be seen if the priors used in this work are broad enough. Such a bias could also impact one's conclusions on fundamental physics; as can be seen from comparing the contours in \cref{fig:priors_cpl_me}, the narrower prior also leads to a better agreement with the $\Lambda$CDM-like late time evolution.

In practice, this explains why analyses that adopt $-1 < w_a < 1$ (as in~\cite{Schoneberg:2024ynd}) find $H_0 \simeq 69$~km\,s$^{-1}$\,Mpc$^{-1}$, whereas with a broader prior the same datasets allow values closer to $H_0 \simeq 65$~km\,s$^{-1}$\,Mpc$^{-1}$. 
Broadening the prior reveals a larger extent of the parameter space, while restricting the prior artificially suppresses the correlated region and biases the apparent constraints on $H_0$.

This exercise demonstrates that prior choices on $w_a$ are not innocuous in CPL-like models once strong degeneracies are present. Analyses adopting $-1 < w_a < 1$ (as in~\cite{Schoneberg:2024ynd}) will inevitably push posteriors toward higher $H_0$ relative to those obtained with broader priors, explaining the apparent discrepancy between our findings and theirs. Moreover, even for our priors, the posterior accumulates at the border of the prior in this example, implying these results may also be suffering from a similar bias, and broadening the prior range further may shift the results.

\end{document}